\documentclass[fleqn,usenatbib]{mnras}

\usepackage{newtxtext,newtxmath}

\usepackage[T1]{fontenc}

\DeclareRobustCommand{\VAN}[3]{#2}
\let\VANthebibliography\thebibliography
\def\thebibliography{\DeclareRobustCommand{\VAN}[3]{##3}\VANthebibliography}

\usepackage{makecell}

\usepackage{graphicx}	
\usepackage{float}
\usepackage{amsmath}	
\usepackage{color}
\usepackage{tabularx}
\usepackage[flushleft]{threeparttable}

\begingroup
\catcode`\_=\active
\gdef_#1{\ensuremath{\sb{\mathrm{#1}}}}
\endgroup
\mathcode`\_=\string"8000
\catcode`\_=12

\def\eg{{\it e.g.}}
\def\ie{{\it i.e.}}

\def\wisk#1{\ifmmode{#1}\else{$#1$}\fi}
\def\msun   {\wisk{{\rm M_\odot}}}
\def\hide#1{}

\title[Stellar Haloes in Dwarf Galaxies]{Ghostly Stellar Haloes and their Relationship to Ultra-faint Dwarfs}

\author[M. Ricotti, E. Polisensky \& E. Cleland]{
Massimo Ricotti$^{1}$\thanks{E-mail: ricotti@umd.edu},
Emil Polisensky$^{2}$,
and Emily Cleland$^{3}$
\\
$^{1}$Department of Astronomy, University of Maryland, College Park, 20742, USA\\
$^{2}$Code 7213, Naval Research Laboratory, Washington, DC, 20375, USA\\
$^{3}$Lockheed Martin, Herndon, VA, 20171, USA
}

\date{Accepted XXX. Received YYY; in original form ZZZ}

\pubyear{2022}

\begin{document}
\label{firstpage}
\pagerange{\pageref{firstpage}--\pageref{lastpage}}
\maketitle

\begin{abstract}
Ghostly stellar haloes are extended haloes of stars composed solely of debris of pre-reionization fossil galaxies and should exist in dwarf galaxies with total masses $<10^{10}$~M$_\odot$. Fossil galaxies are even smaller mass dwarf galaxies that stopped forming stars after the epoch of reionization and have been identified in the Local Group as the ultra-faint dwarf satellites.
Using cosmological N-body simulations we present an  empirical model for the shape and mass of ghostly stellar haloes.
We compare the model to available observations of stellar haloes in six isolated dwarf galaxies in the Local Group (Leo~T, Leo~A, IC~10, WLM, IC~1613, NGC~6822) to infer the star formation efficiency in dwarf galaxies at the epoch of reionization.
We find an efficiency of star formation in dark matter haloes with masses $10^6-10^8$~M$_\odot$ at $z\sim 7$ in rough agreement with independent methods using data on the luminosity function of ultra-faint dwarf galaxies but systematically higher by a factor of 3-5. The systematic uncertainty of our results is still large, mainly because available observations of stellar halo profiles do not extend over a sufficiently large distance from the center of the host dwarf galaxy. Additional observations, easily within reach of current telescopes, can significantly improve the accuracy of this method and can also be used to constrain the present day dark matter masses of dwarf galaxies in the Local Group.
 Our method is based on a set of observations never used before, hence it is a new independent test of models of hierarchical galaxy formation.
\end{abstract}

\begin{keywords}
Local Group -- galaxies: dwarf -- galaxies: high-redshift -- galaxies: haloes -- galaxies: star formation -- reionization
\end{keywords}



\section{Introduction}

Since 2005, the discovery of a new ultra-faint (UF) dwarf
population in data from the Sloan digital sky survey
\citep{ Willmanetal05AJ, Willmanetal05ApJ, Zuckeretal06a, Zuckeretal06b, Belokurovetal06a, Belokurovetal07, Irwinetal07, Walshetal07} and
a survey of M31 \citep{Martinetal06, Ibataetal07, Majewskietal07} has more than doubled the number of known dwarf satellites of the Milky Way and Andromeda.
More recently, searches using data from the Dark Energy Survey \citep[DES;][]{Bechtoletal:2015, KoposovB:15, KimJerjen:2015b,DrlicaWetal:2015,Luqueetal:2016, Nadleretal:2020, DrlicaWetal:2020}, other DECam surveys such as MagLiteS, SMASH, and DELVE
\citep{Martinetal:2015, Drlica-Wagneretal:2016, Torrealbaetal:2018, Koposovetal:2018, Mauetal:2020},  Subaru Hyper Suprime-Cam Survey \citep{Hommaeta:2016,Hommaetal:2018,Hommaetal:2019}, ATLAS \citep{Torrealbaetal:2016a, Torrealbaetal:2016b}, Pan-STARRS1 \citep{Laevensetal:2015a,Laevensetal:2015b}, and Gaia \citep{Torrealbaetal:2019b} have further increased the sample of confirmed and candidate satellites to $\sim 50$.
The existence of a population of true fossils of the first galaxies in the Local Group, with properties similar to UF dwarf galaxies, was first postulated in \cite{RicottiG:05}, just before their observational discovery. The proposal was based on results of cosmological simulations \citep{RicottiGS2002a,RicottiGS2002b} in which positive feedback effects from UV radiation produced a population of faint, but numerous, dwarf galaxies forming in haloes below the Lyman cooling limit (\ie, $M_{\rm halo}<10^8$ M$_\odot$) at $z \sim 6-15$. Later theoretical work confirmed that the newly discovered
UF dwarfs have properties in good agreement with predictions of cosmological simulations of the first galaxies \citep{BovillR2009,BovillR2011a,BovillR2011b,RicottiPG2016}. More recently, zoom simulations of dwarf galaxies and the satellites of the Milky Way have arrived at similar conclusions \citep{Wheeleretal:2015, Wetzeletal:2016, Munshi:2017, Wheeleretal:2019, Munshietal:2019, Garrison-Kimmeletal:2019,Agertzetal:2020, Samueletal:2020}. The identification of UF dwarfs as "fossil" galaxies is confirmed by data on their star formation histories \citep{Grebel:04, Brown:12, Brown:14}.

If the identification of UF dwarf galaxies as fossils is correct, we expect a large population to be still undetected but accessible to future deep surveys such as the Rubin Observatory Large Synoptic Survey Telescope (LSST). Another notable consequence is the existence of faint stellar haloes around isolated dwarf galaxies (which have not been tidally stripped by the Milky Way or Andromeda). Stellar haloes are extended and faint stellar structures formed by debris of tidally disrupted dwarf galaxies accreted over time by the host galaxy. Around dwarf galaxies, these stellar haloes may not exist if all the accreted satellites are dark haloes without stars. However, if a stellar halo is found in sufficiently small mass dwarfs, the whole stellar halo can be composed of tidal debris of UF fossil galaxies. Such hypothesised stellar haloes have been referred to as "ghostly stellar haloes" \citep{BovillR2011a} and their properties can be directly related to the properties of the UF dwarf population. 

Stellar haloes have been observed in the outskirts of Local Group dwarfs and beyond \citep[\eg,][]{Higgsetal:2016, Kado-Fongetal:2020,Higgsetal:2021}. \citet{KangR:2019} collected data in the literature for six Local Group dwarf galaxies showing evidence of an extended stellar component: Leo~T \citep{LeoT}, Leo~A \citep{LeoA, LeoADwarf}, WLM \citep{WLMFringe, WLMKeck, Minniti}, IC~1613 \citep{IC1613, IC1613AGB,Puchaetal:2019}, IC~10 \citep{IC10}, and NGC~6822 \citep{NGC6822}. \citet{KangR:2019} interpreted the extended stellar component as ghostly stellar haloes and using a simple semi-analytic model, showed that dark matter haloes in the mass range $10^6-10^8$~M$_\odot$ at $z\sim 6-7$ have a mean star formation efficiency in the range $f_* \equiv M_*/M_{dm} \sim 0.1\%-0.2\%$, only mildly increasing as a function of the dark matter halo mass. This result extends to lower halo masses previous published works on the star formation efficiency in galaxy haloes with $M_{\rm dm}>10^{10}$~M$_\odot$ \citep{Behroozi2013}.

The main limitation in the early work by \cite{KangR:2019} is the analytic derivation of the stellar halo model that relies on a few assumptions, namely the homology of the stellar profiles and other simplifications.
In this paper we improve on the work of \citet{KangR:2019} by formulating a more accurate ghostly stellar halo model based on a set of N-body simulations. The model aims at characterizing the sizes and stellar masses of ghostly stellar haloes in dwarf galaxies as a function of their dark matter halo mass and the efficiency of star formation in the first galaxies. Using this more accurate model we provide updated predictions on the star formation efficiency in small mass dark matter haloes before the epoch of reionization ($M_{dm}<10^8$~M$_\odot$ at $z\sim 6-7$), using the same observational data collected in \cite{KangR:2019}.

This paper is organized as follows. In \S~\ref{sec:sim} we present the numerical methods and details about the N-body simulations. The results of this work are presented in \S~\ref{sec:results}, including the formulation of a physically motivated model of stellar haloes, calibrated to reproduce the result of a grid of N-body simulations. In \S~\ref{sec:constraints} we apply the empirical model to observations of extended stellar haloes in six isolated dwarfs in the Local Group and derive constraints on $f_*$ at the epoch of reionization. In \S~\ref{sec:disc} we discuss present and future applications of this method and compare it to previous results from the literature. Summary and conclusions are in \S~\ref{sec:conc}.

\section{Methodology and Simulations}\label{sec:sim}

We use the `zoom-in' technique of nested refinement volumes inside a larger cosmological environment sampled at lower resolution to simulate formation histories of individual dwarf-scale haloes with masses $\sim 10^{9}$~\msun\ at high resolution.

We adopt cosmological parameters that closely match several large cosmological simulations, including Bolshoi \citep{bolshoi} and the Juropa Hubble Volume simulation \citep{jub2014}: $(\Omega_m, \Omega_\Lambda, \Omega_b, h, \sigma_8, n_s)$ = $(0.27, 0.73, 0.044, 0.7, 0.8, 0.96)$.
Initial conditions are generated with {\sevensize MUSIC} \citep{music} using second-order Lagrangian perturbation theory with power spectrum and transfer functions calculated by {\sevensize CAMB} \citep{camb}. The $N$-body cosmological simulation code {\sevensize GADGET} \citep{spr2005} is used to evolve the simulations and {\sevensize AHF} \citep{ahf} for finding haloes composed of more than 20 gravitationally bound particles. We use an overdensity criterion of 178 times the critical density for defining the halo virial radius and mass at all redshifts. The {\sevensize MergerTree} tool is used for determining halo progenitors and constructing merger trees.

The starting redshift of all simulations is set to when the rms variance of matter distribution is 0.13. Dark matter force softening lengths are set to 1.5\% the grid cell side length and held constant in comoving units for $z>9$ and constant in physical units subsequently.

\subsection{Dark Matter Simulations}

\begin{figure*}
\centering
\includegraphics[width=1.0\textwidth]{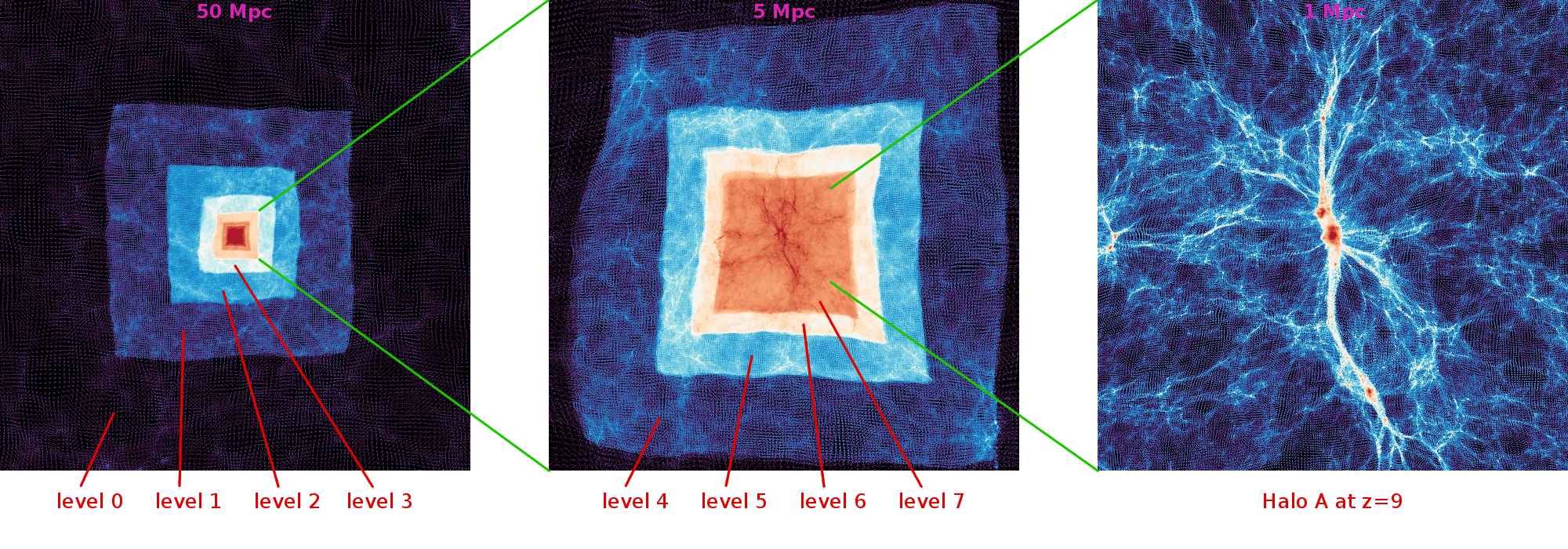}
\caption{Large scale structure of the dark matter and nested zoom layers at different resolution for Halo~A at $z=9$. We use seven levels of refinement to evolve the haloes in Table~\ref{tab:sims} to redshift $z=0$ and achieve a dark matter mass resolution of 100~$\msun$ in haloes with mass $< 10^9$~$\msun$ and 1000~$\msun$ in haloes more massive than $10^9$~$\msun$. The mass resolution of the stars is 100 times smaller than the dark matter particle mass (\ie, 1~$\msun$ and 10~$\msun$).\label{fig:zoom}}
\end{figure*}

We adopt a fiducial comoving cubic simulation volume 49.3 Mpc on a side containing up to seven refinement volumes. We refer to the coarsest mass resolution covering the entire cube as ``level 0'' and sequentially label the nested, increasing resolution zoom levels. Dark matter simulation particle mass varies a factor of eight between levels reaching a minimum of $m_{dm} = 1000$~\msun\ in level 7. In Fig.~\ref{fig:zoom} we show a slice through the simulation volume at $z=9$ to illustrate the multilayered setup.

\begin{figure}
\centering
\includegraphics[width=0.4\textwidth,angle=270]{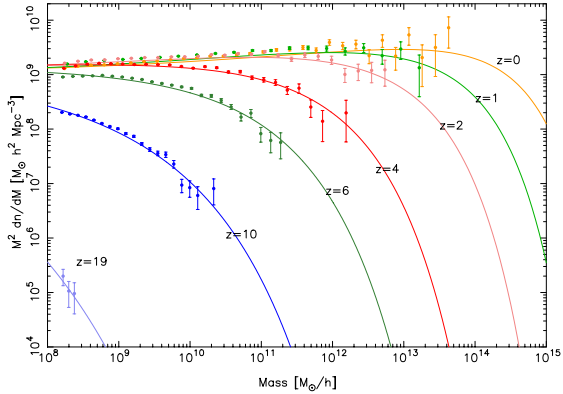}\\
\caption{Mass functions of level 3 unigrid simulation at several redshifts. Solid lines are the analytic functions of \citet{Behroozi2013}.\label{fig:mf}}
\end{figure}

We begin by generating unigrid initial conditions where the level 3 volume covers the entire simulation cube: $1024^3$ particles with mass resolution $4.096 \times 10^6$~\msun. The initial redshift is $z_i=65$ and the force softening length is 722~pc. Fig.~\ref{fig:mf} shows the simulation halo mass function at several redshifts. Model mass functions are calculated with {\sevensize HMF}\footnote{Predecessor to {\sevensize TheHaloMod} \citep{thm2021}} \citep{hmf2013}. At all redshifts our simulation shows good agreement with fitting functions derived from the Bolshoi simulation \citep{Behroozi2013}, as expected for our adopted cosmological parameters.

\begin{figure*}
\centering
\includegraphics[width=0.9\textwidth]{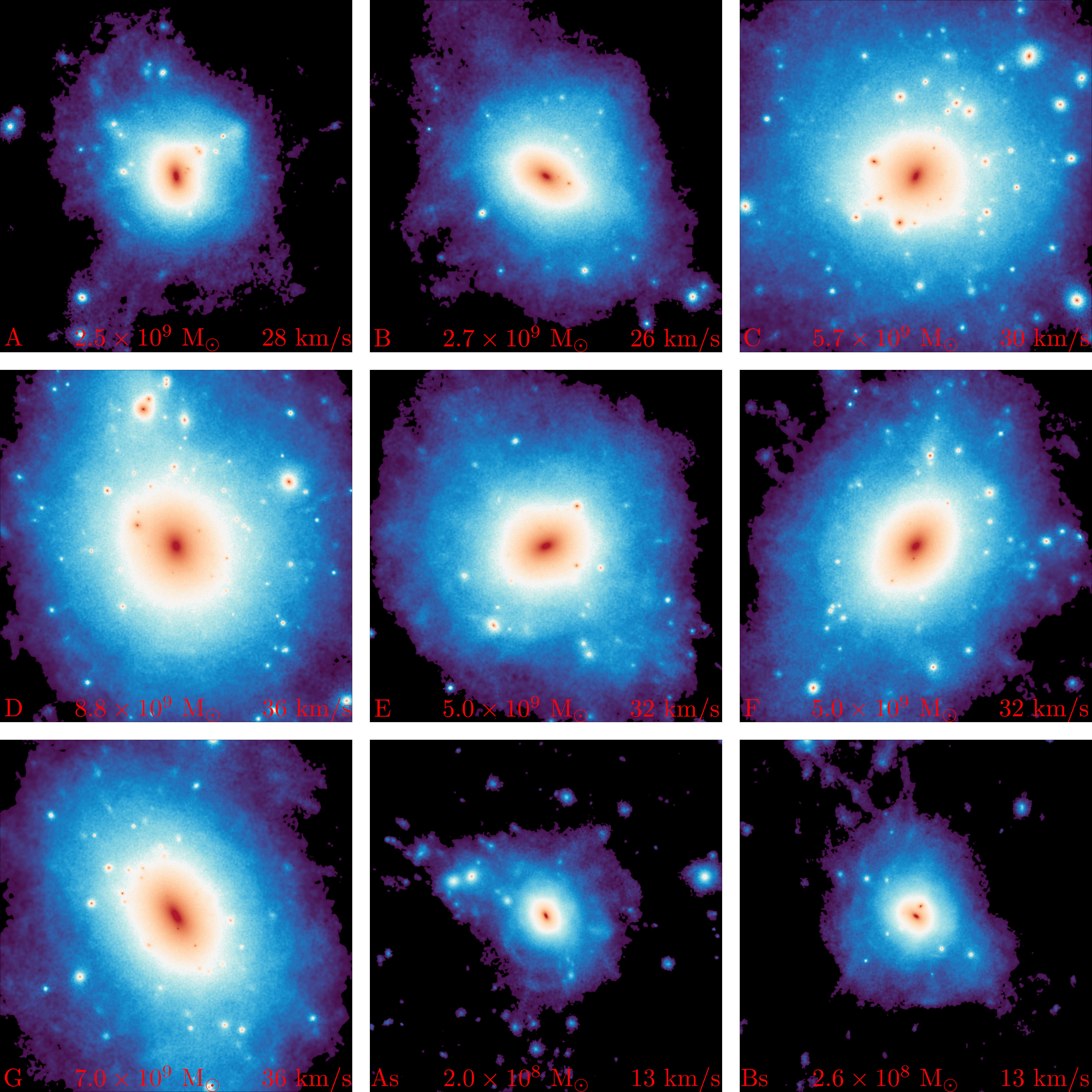}
\caption{Highest resolution dark matter portraits at $z=0$ of our isolated dwarf galaxies. Simulation particles within a (100 kpc)$^3$ volume centered on each halo are rendered. Virial mass and maximum circular velocity for each halo are labeled.}\label{fig:dm_portraits}
\end{figure*}

We examined the growth histories of haloes with mass $10^9-10^{10}$~\msun\ at $z=0$ and inspected their locations in the simulation volume at all redshifts. We selected seven that appeared isolated, had identifiable progenitors for at least $z<6$, had not fallen into larger structures or merged with a halo $> 50\%$ its mass for $z < 6$. Our selected haloes, lettered A-G, have masses ranging $2.5 - 8.8 \times 10^9$~\msun\ and maximum circular velocities $v_c = 26-36$ km/s at $z=0$, where $v_c^2 = GM(r)/r$ and $M(r)$ is the mass enclosed within radius $r$. Fig.~\ref{fig:dm_portraits} shows dark matter portraits at $z=0$ and gives the virial mass and maximum circular velocity for each halo.

\begin{figure}
\centering
\includegraphics[width=0.35\textwidth,angle=270]{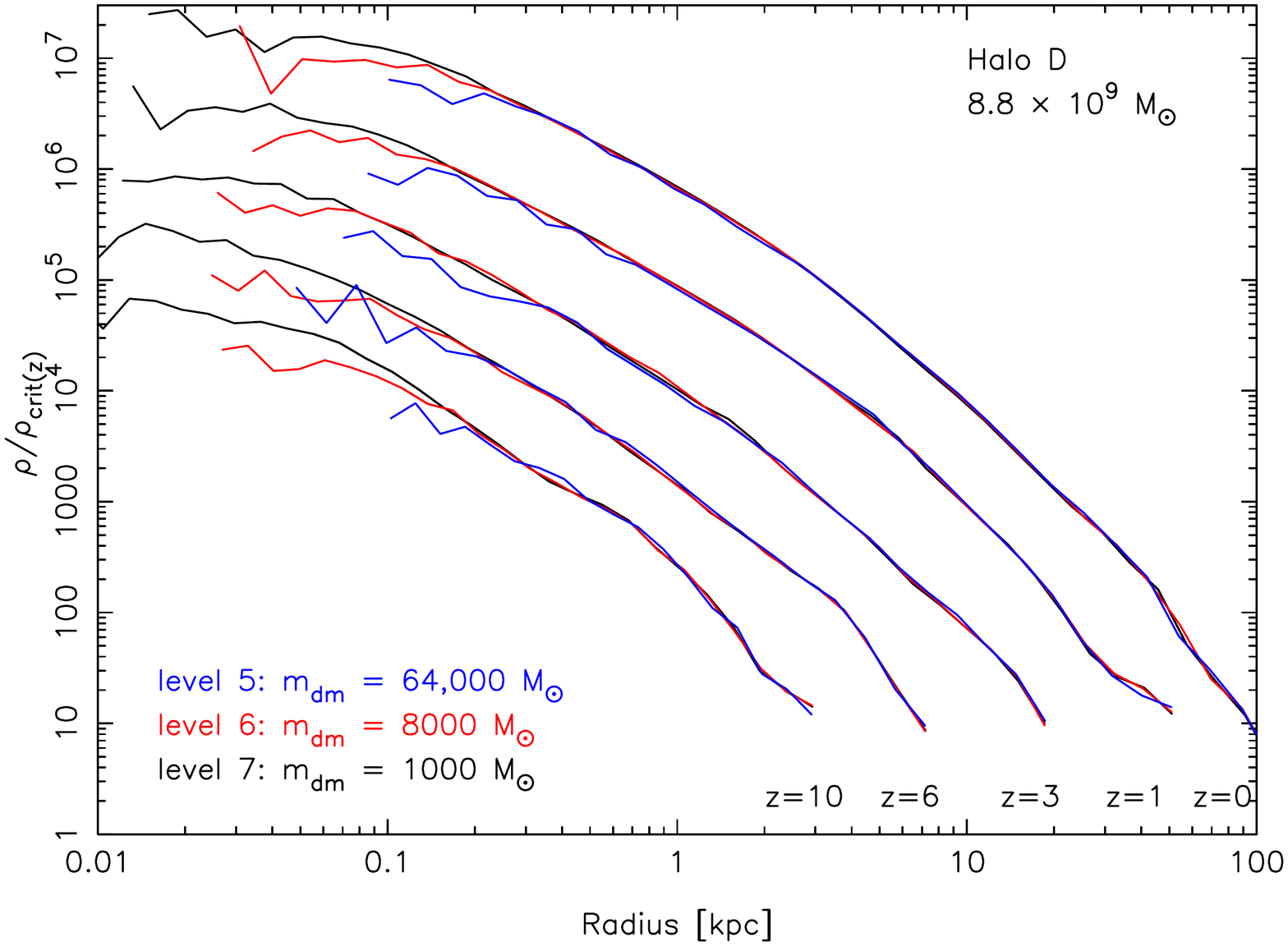}\\
\caption{Density profiles of Halo D at several redshifts (z = 10, 6, 3, 1, 0) in the three highest mass resolution simulations. Profiles are normalized by the critical density at the time of each redshift.\label{fig:den}}
\end{figure}

Each halo was simulated at each higher resolution from level 4-7. Refinement volume placement and extent were determined by tracing $z=0$ halo particles to the initial conditions. Level 7 volumes are approximately 1.5~Mpc on a side. The density field in the coarser resolution levels (0-2) was created by averaging down the level 3 initial conditions with {\sevensize MUSIC}. For the highest resolution simulations with all seven refinement levels, $z_i = 110$ and the softening length is 45~pc. Fig.~\ref{fig:den} compares density profiles of our largest halo at several redshifts across the three highest mass resolution. Differences in the inner profile due to the resolution dependent softening length are apparent but otherwise our simulations show good agreement at all times and across mass resolutions, validating our methodology.

To sample dwarf haloes $<10^9$~\msun, the simulation box for Haloes A and B was shrunk to 22.9 Mpc on a side, reducing the halo masses an order of magnitude and the simulation mass resolution to 100~\msun. Seven levels of refinement were used and the starting redshift and force softening length scaled appropriately ($z_i = 127$, softening length 21~pc). To indicate their smaller size we label these haloes As and Bs where their $z=0$ masses are $\sim 2 \times 10^8$~\msun\ and maximum circular velocities $13$~km/s.

\subsection{Simulations with Dark Matter and Stars}

With the seven level, dark matter only simulations in hand we add star particles representing the fossil stellar populations and evolve the simulations from the end of the epoch of reionization to $z=0$. We take $z=7$ as the time when star formation in fossil galaxies is quenched by the heating of the IGM. Progenitor haloes at $z=7$ that contribute mass to the halo at $z=0$ are populated with star particles using a model for the amount and distribution of stellar mass.

We do not expect quenching of star formation due to reionization to be instantaneous. In \cite{KangR:2019} we explored, using semi-analytic models, the effect of delaying the reionization quenching of star formation from $z_{rei}=6$ to $z_{rei}=5$. We found a negligible effect on the scale radius of the stellar halo and a small effect on the total stellar halo mass. Therefore a more realistic model of reionization, without a sharp cutoff, should have a small impact on our results. Clearly the constraints on $f_*(z)$ have to be interpreted to refer to the effective quenching redshift $z=z_{rei}$.

Motivated by semi-analytic models of star formation during the epoch of reionization \citep{hart2016}, the halo stellar masses are modeled assuming the following simple power law relationship for the mean star formation efficiency:
\begin{equation}
f_*(M) \equiv 
\begin{cases}
\epsilon_0(M/M_0)^\beta, & M>M_{cut},\\
0, & M<M_{cut}. 
\end{cases} \label{eq:fstar}
\end{equation}
where $\epsilon_0$, $\beta$ and $M_{cut}$ are free parameters. 

The sky-projected two dimensional surface brightness profiles of dwarf galaxies are well described by a S\'ersic law:
\begin{equation}
    \Sigma(R) = \Sigma_0 \exp \left(-(R/R_0)^{1/n}\right)
\end{equation}
where $R$ is the two dimensional radial coordinate on the sky and $R_0$ is a scale length.
\citet{neto99} give an analytic approximation for the deprojected, three dimensional enclosed stellar mass profile:
\begin{equation}
    \frac{M(r)}{M_*} = \frac{\gamma(n(3-p),(r/R_0)^{1/n})}{\Gamma(n(3-p))}
\label{eq:massratio}
\end{equation}
where $\Gamma(b)$ is the complete gamma function and $\gamma(b,x)$ is the lower incomplete gamma function. \citet{marq00} give an updated value of the fitting parameter, $p = 1.0 - 0.6097/n + 0.05563/n^2$. 
In this study we only consider exponential disk profiles where $n=1$ and the ratio of the scale length to the three-dimensional half-mass radius is: $R_0/r_{eff} = 0.4483$. We define a free parameter for the ratio of half-mass to virial radius, $\eta \equiv r_{eff}/r_{vir}$, to relate the scale length to the halo virial radius determined from the dark matter simulations.

Star particle velocities are modeled assuming they are in virial equilibrium within the potential well of the dark matter halo. The stellar velocity dispersion in each dimension is
\begin{equation}
    \sigma^2 = GM_{eff}/3r_{eff}
\label{eq:sigma}
\end{equation}
where $M_{eff}$ is the total mass within $r_{eff}$. $M_{eff}$ is calculated from the dark matter particles using the value of $r_{eff}$ scaled from the virial radius by the adopted value of $\eta$. 

Equations~(\ref{eq:fstar}), ~(\ref{eq:massratio}) and~(\ref{eq:sigma}) define our model for the fossil stellar populations. For each dark matter halo, we determine the number of star particles from the model total stellar mass [equation~(\ref{eq:fstar})] and adopted star particle mass resolution, $m_*$. We keep the ratio of simulation dark matter to star particle mass resolution constant and adopt $m_* = 10$~\msun\ for Haloes A-G and $m_* = 1$~\msun\ for Haloes As and Bs. We use an inverse transform technique to randomly generate radial coordinates for star particles following the mass distribution. A sample from a random variable uniformly distributed between 0 and 1 is generated. The three dimensional radius originating at the halo center of mass is calculated by numerically solving for where the enclosed stellar mass ratio [equation~(\ref{eq:massratio})] equals the random number to within a small tolerance ($10^{-8}$). Two more uniform random variates, $u_1, u_2$, are used to calculate the spherical angles: $\theta = 2\pi u_1$, $\phi = \arccos(2u_2 -1)$. Star particle coordinates are converted from spherical to Cartesian and the halo center of mass coordinates added.

Particle velocities in each dimension are assigned by multiplying $\sigma$ [equation~(\ref{eq:sigma})] with a normally distributed random variable of zero mean and unit standard deviation. The halo center of mass velocity is calculated from dark matter particles within $3r_{eff}$ and added to the star particle velocities.

Our model of the fossil stellar populations has 4 independent parameters: $M_{cut}$, $\beta$ and $\epsilon_0$ that determine the total stellar mass; and $\eta$ that determines the scale length of the stellar distribution. We set the pivot point $M_0 = 10^8$~\msun\ in equation~(\ref{eq:fstar}) and hold it constant for all simulations. \citet{krav2013} finds $\eta \approx 0.015$ for galaxies over eight orders of stellar mass and all morphological types at the present time. We adopt $\eta=0.15$ for haloes during the epoch of reionization assuming the dependence on scale factor is dominated by the linear scaling of the virial radius. However, we also tested the sensitivity of the results to other choices of $\eta$ (see \S~\ref{ssec:reff}).

The properties of the stellar halo are trivially dependent on the choice of $\epsilon_0$, as the surface brightness of the halo profile depends linearly on $\epsilon_0$. We therefore set $\epsilon_0 = 0.003$ and do not explore further the dependence of the simulation results on this parameter. We instead run a grid of simulations with different values of $\beta$ (typically $\beta=-0.5, 0, 0.5, 1$) for $M_{cut}=10^6$~\msun\ and for $M_{cut}=10^7$~\msun.

We decided to use only a subset of our haloes to explore the dependence of the stellar halo profile covering the widest range of mass while allowing more trials of the model free parameters. We simulated four main haloes with the following $z=0$ total mass: Halo~As ($2.0\times 10^8$~\msun), Halo~Bs ($2.6 \times 10^8$~\msun), Halo~B ($2.7 \times 10^9$~\msun), and Halo~D ($8.8 \times 10^9$~\msun). We also partially explore the mass variance by running a trial with Halo~G ($7.0 \times 10^9$~\msun), which has a similar mass to Halo D (see \S~\ref{ssec:reff}).
 
Table~\ref{tab:sims} presents the complete set of 28 simulations with dark matter and star particles we evolved to $z=0$. We adopt star particle softening lengths of 8~pc for haloes A-G and 1.72~pc for As and Bs. In a similar manner we also reduced the level 7 dark matter particle softening length to 21.6~pc and 10~pc for these simulations.

\begin{table*}
	\centering
	\caption{Table of the set of 28 simulations with dark matter and star particles.}
	\label{tab:sims}
	\begin{threeparttable}
	\begin{tabular}{l|cccccccc} 
		\hline
		Halo ID & $M_{vir}$ [$\msun$] & $\eta$ & $\log{(M_{cut}/\msun)}$ & $\beta$ values & $m_{dm}$ [$\msun$] & $m_*$ [$\msun$] &\multicolumn{2}{c}{Max softening length [pc]\tnote{a}} \\
		& & & & & & & darm matter & stars\\
		\hline
		Halo As & $2.0 \times 10^8$ & 0.15 & 6     & -0.5, 0, 0.5, 1 & 100 & 1 & 10 & 1.72\\
		Halo Bs & $2.6\times 10^8$ & 0.15 & 6     & -0.5, 0, 0.5, 1 & 100 & 1 & 10 & 1.72\\
		Halo B6 & $2.7\times 10^9$ & 0.15 & 6 & -0.5, 0, 0.5, 1 & 1000 & 10 & 21.6 & 8\\
		Halo B7 & $2.7\times 10^9$ & 0.15 & 7 & -0.5, 0, 0.5, 1 & 1000 & 10 & 21.6 & 8\\
		Halo D6 & $8.8\times 10^9$ & 0.15 & 6 & -0.5, 0, 0.3, 1 & 1000 & 10 & 21.6 & 8\\
		Halo D7 & $8.8\times 10^9$ & 0.15 & 7 & -0.5, 0, 0.7, 1 & 1000& 10 & 21.6 & 8\\
        Halo D7r & $8.8\times 10^9$ & \makecell{0.025, 0.08, 0.3} & 7 & 0 & 1000 & 10 & 21.6 & 8\\
		Halo G & $7.0\times 10^9$ & 0.15 & 6     & 0 & 1000 & 10 & 21.6 & 8\\
		\hline
	\end{tabular}
	\begin{tablenotes}
\item[] 
(a) The softening length is given in physical units at $z=9$.
Dark matter force softening
lengths are set to 1.5\% the grid cell side length and held constant
in comoving units for $z>9$ and constant in physical units subsequently.
Similar scaling applies to the force softening of the stars.
\end{tablenotes}
\end{threeparttable}
\end{table*}

\section{Simulation Results and Modelling of Stellar Halo Profiles}\label{sec:results}

In this section we present the analysis of the simulations results, starting with the evolution of the dark matter halo profiles and the merger histories of our isolated dwarf galaxies (\S~\ref{ssec:merger}). Next we study the spherically averaged density profiles of the stars in each halo (\S~\ref{ssec:profiles}), and based on the simulation results we propose an empirical, but physically-motivated, model reproducing the profiles of the stellar haloes extracted from the simulations (\S~\ref{ssec:model}).

\subsection{Evolution of the Dark Matter Profile and Merger History}
\label{ssec:merger}

\begin{figure*}
\centering
\includegraphics[width=0.49\textwidth]{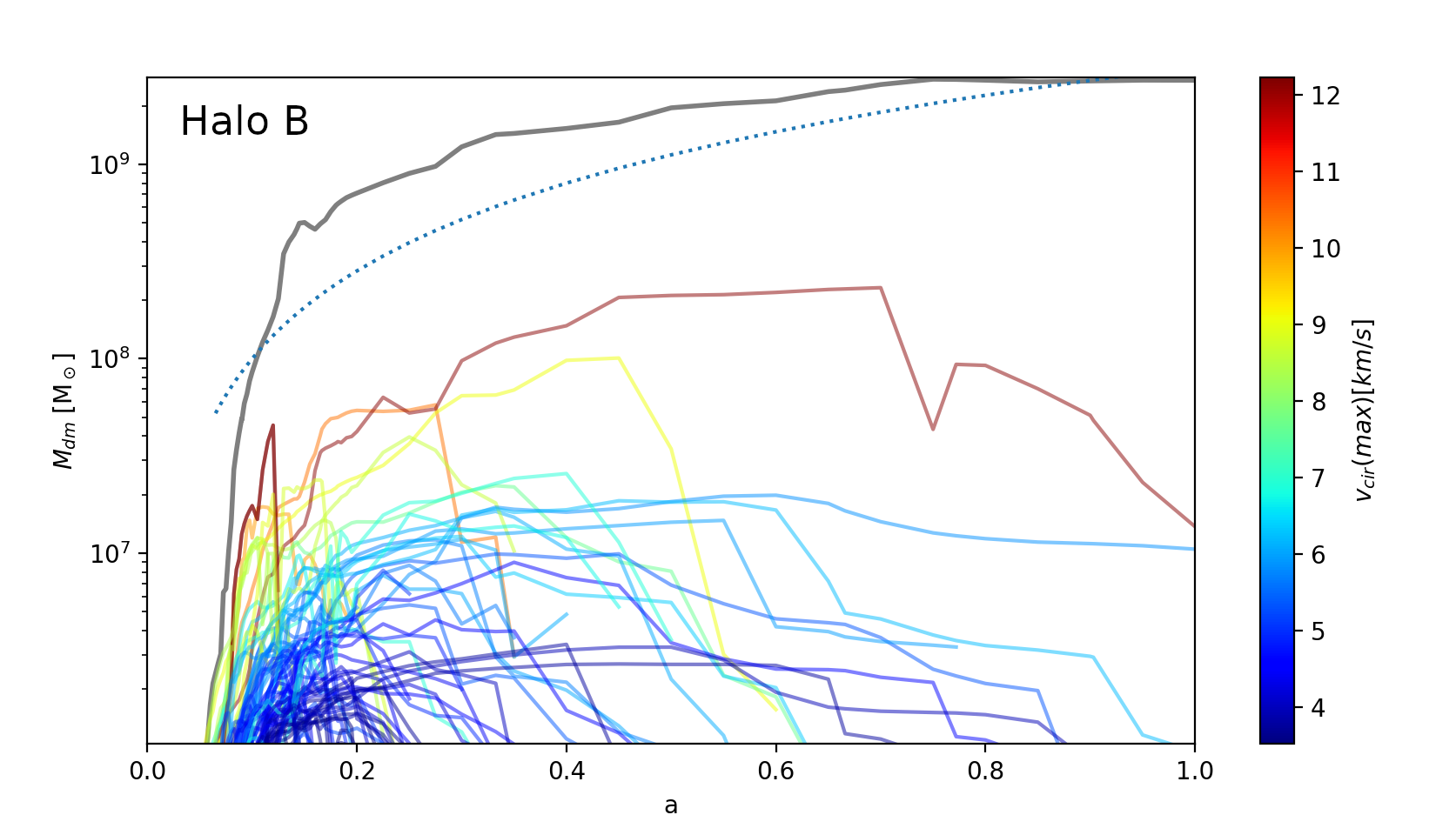}
\includegraphics[width=0.49\textwidth]{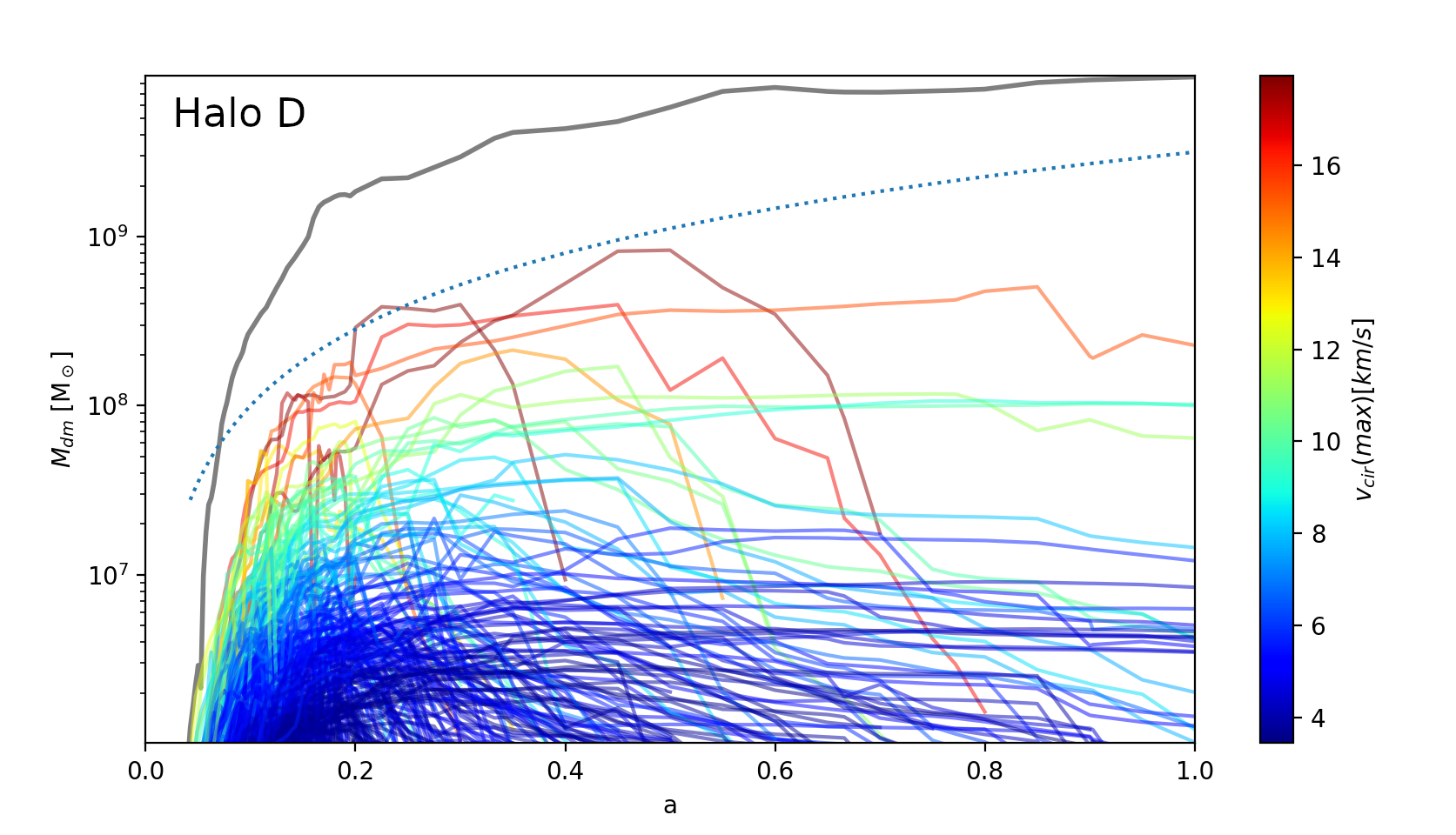}
\caption{Growth and merger histories of Halo~B (left) and Halo~D (right). The thick solid black lines show the mass of the main (central) halo as a function of the scale parameter $a=1/(1+z)$. The colored lines show the halo mass of the satellites of the main halo (haloes that are or have been within the virial radius of the halo), colored according to their maximum circular velocity, as shown by the colorbar. The dotted blue line shows the halo mass, $M_{Jeans,IGM}$, corresponding to virial temperature $T_{vir}=10^4$~K, which is roughly considered the critical mass for atomic cooling haloes and/or haloes in which star formation can be suppressed after reionization of the IGM at $z_{rei} \sim 6-7$. The main Halo~B and Halo~D have masses $>M_{Jeans,IGM}$, hence they are expected to continue forming stars after reionization and the central galaxy will grow mostly after the epoch of reionization. On the contrary, the satellites of all the haloes considered in this study (with $M_{halo}< 10^{10}$~M$_\odot$), have maximum masses that never exceed $M_{jeans,IGM}$, hence these satellite haloes form all their stars before the epoch of reionization in haloes that are below the atomic cooling critical mass. We typically refer to these dwarf galaxies as pre-reionization fossil galaxies, which have been identified with the UF dwarfs in the Local Group.}
\label{fig:mergerhist}
\end{figure*}

We start by analyzing the evolution of the dark matter properties of the isolated haloes in our sample from the time of formation to $z=0$.
 All the simulations share a similar qualitative evolution illustrated in Fig.~\ref{fig:mergerhist}, showing the mass growth histories of all sub-haloes in Halo~B (left) and Halo~D (right). The thick solid black line shows the mass of the main central halo as a function of the scale parameter $a=(1+z)^{-1}$. The colored solid lines shown the evolution of the mass of the satellites. The colors refer to the circular velocity of the satellites as shown in the colorbar. The dotted blue line shows the evolution of the Jeans mass of the IGM assuming $T_{IGM}=10^4$~K after reionization, to emphasize that in these two haloes the satellites have masses that remain always below the Jeans mass in the IGM. Hence, gas accretion and star formation in the haloes merging with the central host is sterilized after reionization \citep{Efstathiou:92b,Gnedin00b} (but see \citet[][]{Ricotti2009, Jeonetal:2017, Reyetal:2019, Reyetal:2020} for models and numerical studies in which late-time -- after reionization -- gas accretion can take place even in UF dwarf galaxies). 
 
 The merger histories shown in Fig.~\ref{fig:mergerhist} motivate our assumption that star formation in sub-haloes merging and contributing to the stellar haloes of the dwarf galaxies in the mass range simulated in this study, can only take place before the epoch of reionization. Therefore the stellar haloes in these dwarfs are dominated by stars forming at redshifts $z>6-7$ (unless the stellar halo is contaminated by stars belonging to the central galaxy that instead can continue to form stars after reionization). 
 Fig.~\ref{fig:mergerhist} also shows that in these small mass haloes evolving in relative isolation after the epoch of reionization, the stellar haloes and the dark matter haloes grow over time due to minor mergers ({\it i.e}, satellite to host mass ratios of $1:10$ or smaller).
 
\begin{figure}
\centering
\includegraphics[width=0.35\textwidth,angle=270]{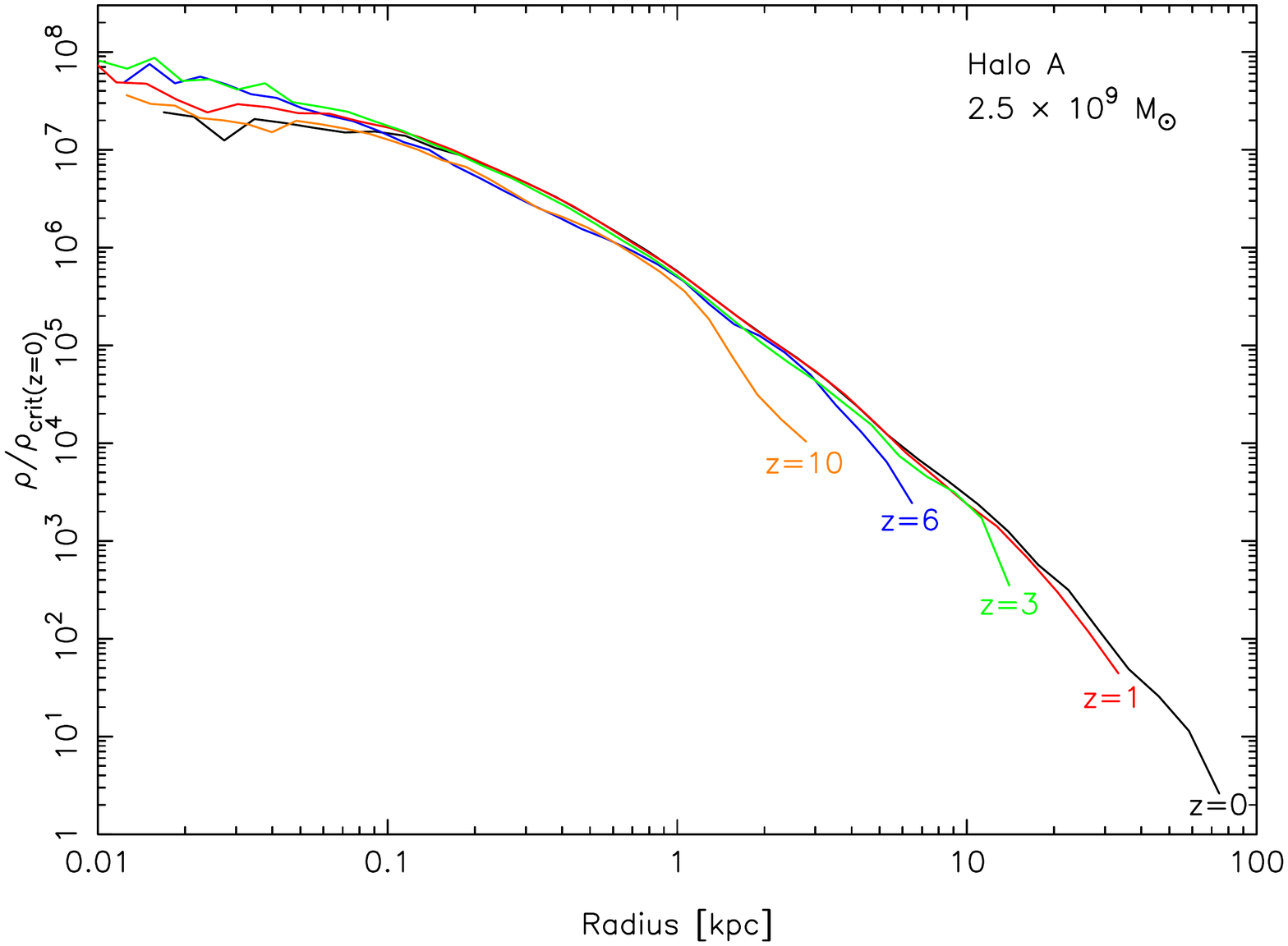}
\caption{Evolution as a function of redshift (at $z=10, 6, 3, 1, 0$) of the density profile of Halo~A at level 7 resolution (dark matter particle mass 1000~M$_\odot$). Profiles are normalized by the critical density at $z=0$. This halo is representative of the typical evolution of the density profile of an isolated small mass halo: the shape of the inner parts of the density profile is set at high-redshift, while the outer parts of the halo profile are assembled at later time when the total mass and virial radius increase as a result of the expansion of the universe and the decreasing mean density of the IGM.}
\label{fig:halogrowth}
\end{figure}

Fig.~\ref{fig:halogrowth} shows the evolution of the dark matter density profile of Halo~A in units of the critical density at $z=0$. The shape of the inner density profile is set at high-redshift, while the outer density profile is assembled later as the halo radius increases with time due to the expansion of the universe (\ie, the decreasing mean density of the universe). Hence, the scale radius $r_s$ of the halo remains constant, while the virial radius $r_{vir}$ and the concentration of the halo $c \equiv r_{vir}/r_s$ both increase with time. This evolution of the density profile can be understood qualitatively in terms of "cosmological secondary infall", and has been explored in detail in several previous studies \citep{Bertschinger:85,Bullock00,Ricotti:03,Ricotti2009,DiemerK2013,DiemerK2014}. 

\subsection{Ghostly Stellar Haloes and their Density Profiles}
\label{ssec:profiles}

\begin{figure*}
\centering
\includegraphics[width=0.33\textwidth]{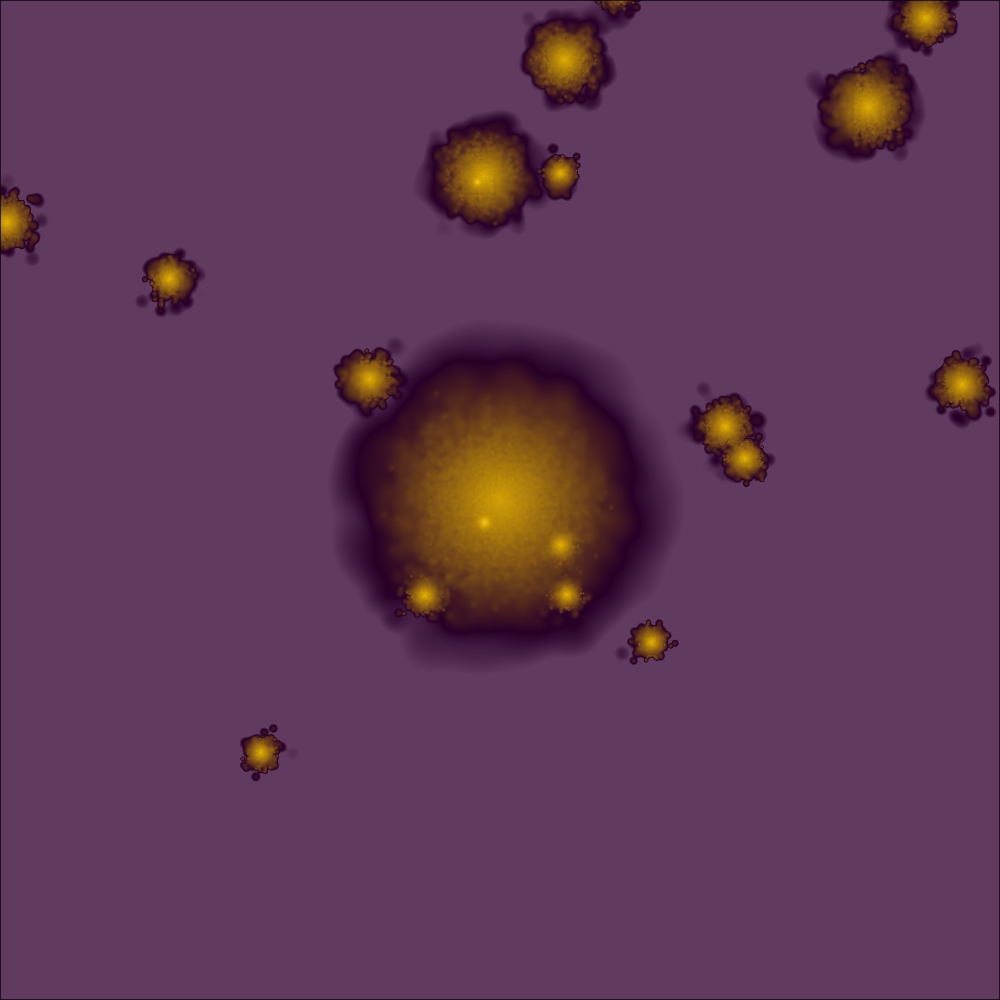}
\includegraphics[width=0.33\textwidth]{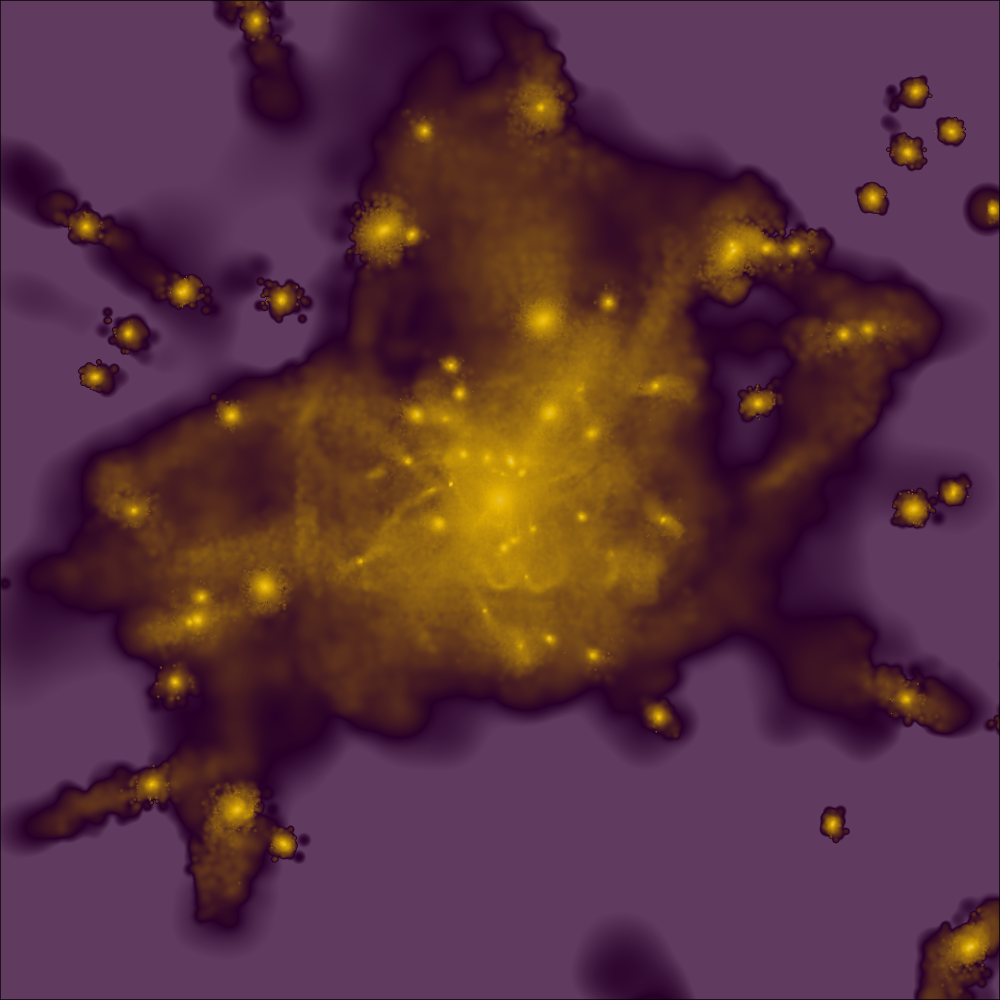}
\includegraphics[width=0.33\textwidth]{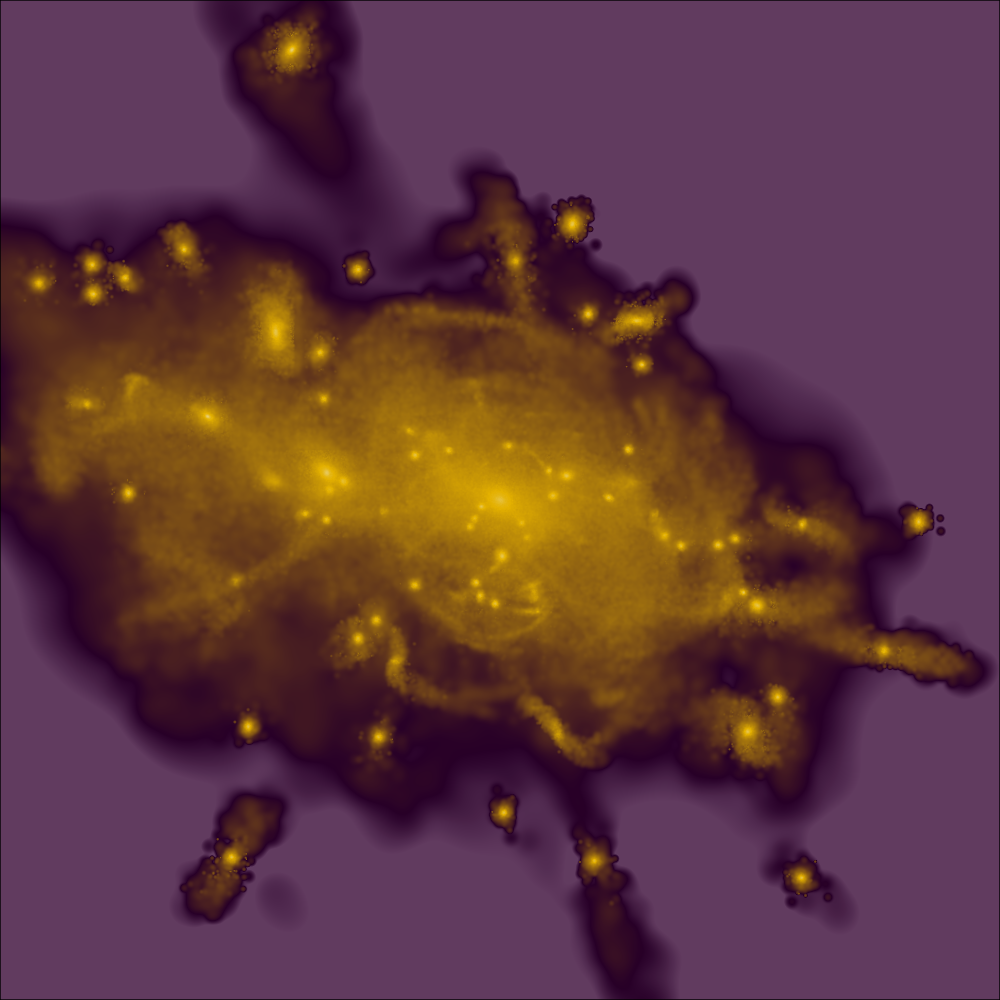}\\
\includegraphics[width=0.33\textwidth]{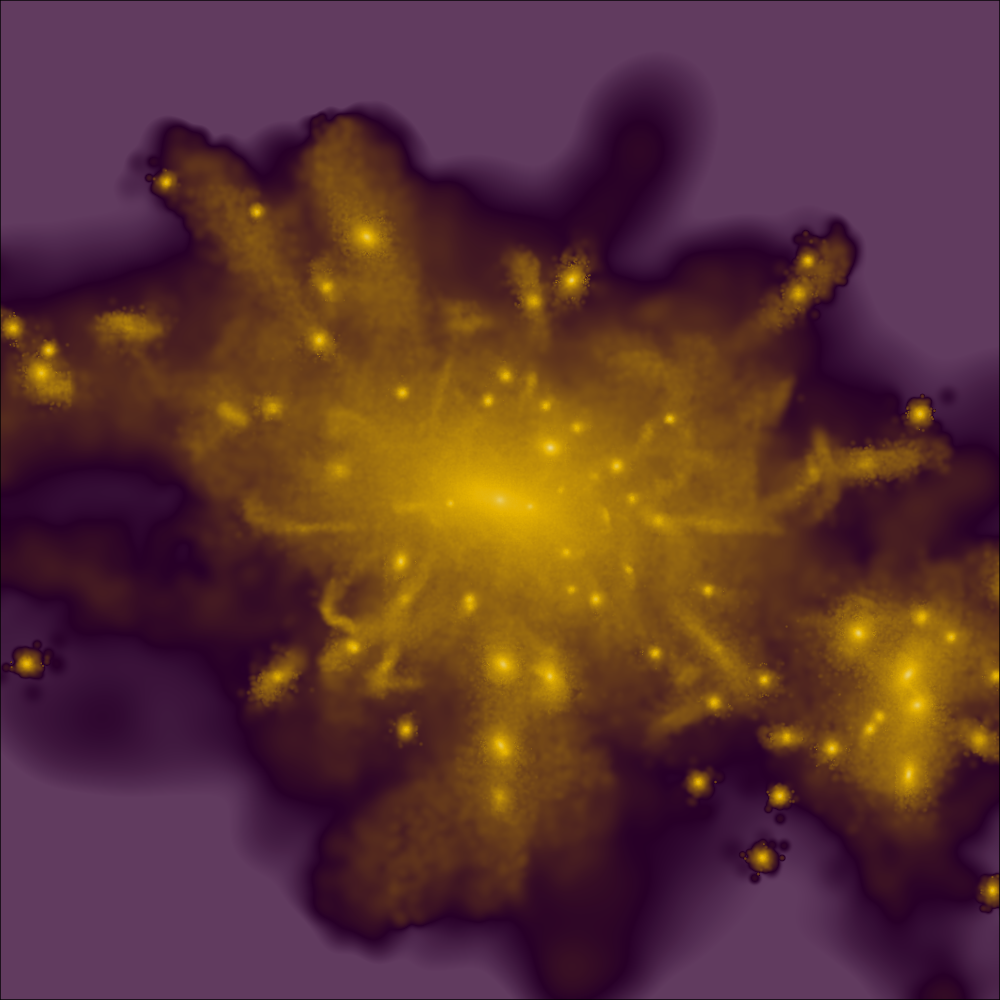}
\includegraphics[width=0.33\textwidth]{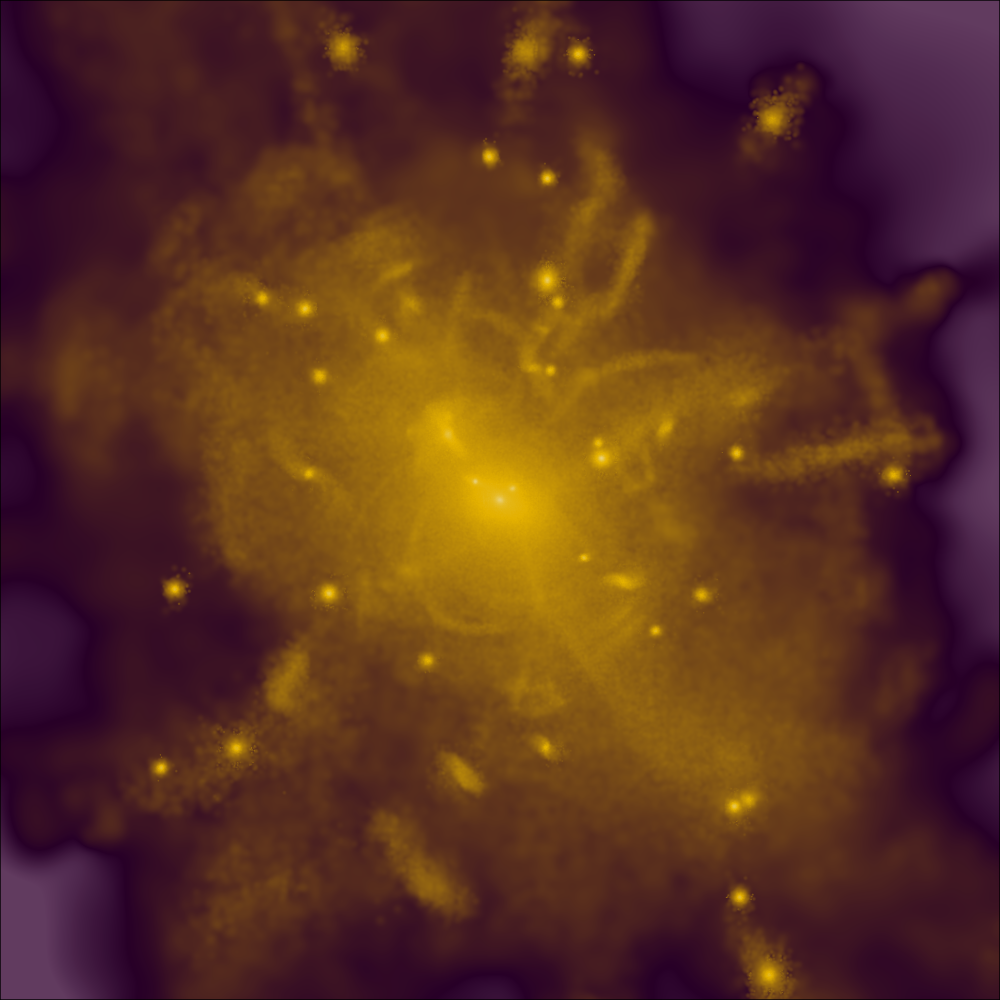}
\includegraphics[width=0.33\textwidth]{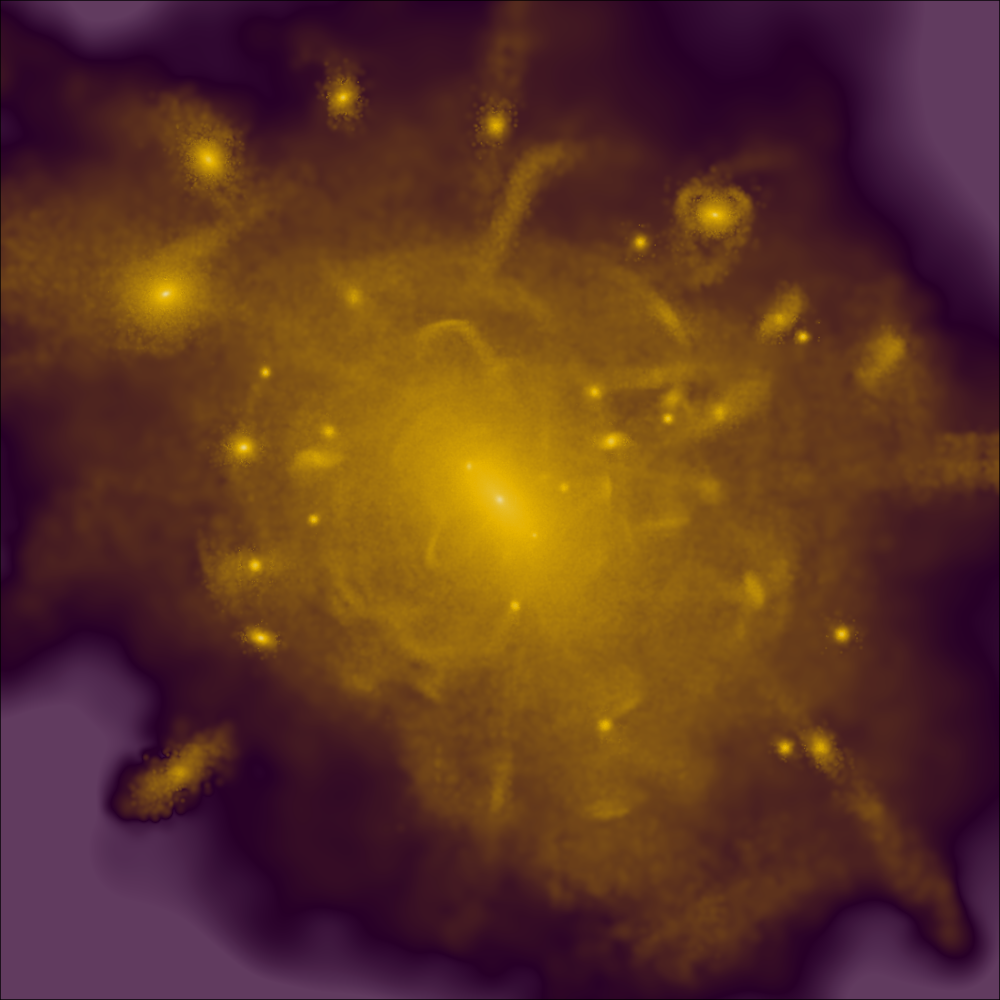}
\caption{Snapshots from a movie of Halo~D with $\beta=-0.5$ and $M_{cut}=10^6$~M$_\odot$, rendering stars (with masses of 10~M$_\odot$) in a (100 kpc)$^3$ comoving volume. Each panel from left to right refers to redshifts $z=7, 2.8, 1.8, 0.9, 0.4, 0$. The first snapshot, at $z=7$, shows the initial conditions in which each dark matter halo is populated with a stellar spheroid with an exponential profile and scale radius $r_{eff}=0.15r_{vir}$. The other snapshots show the accretion and tidal stripping of satellites including some that survive to $z=0$ and others that contribute to the growth of the smoother stellar halo component.}
\label{fig:images}
\end{figure*}

Snapshots showing the evolution of the surface density of stars in Halo~D is shown in Fig.~\ref{fig:images}. The images highlight the presence of satellites, their destruction by tidal effects, as well as the growth and triaxiality of the underlying smoother stellar halo produced by the debris of stripped satellite galaxies. The six snapshots shown in the figure refer to run Halo D6 in Table~\ref{tab:sims}, with total mass at $z=0$ $M_{halo}=8.8 \times 10^{9}$~\msun, $\beta=-0.5$, $M_{cut}=10^6$~\msun\ and mass of dark matter particles and stars of $1000$~\msun\ and $10$~\msun, respectively. The negative value of $\beta$ for this simulation is such that small mass haloes with masses $>10^6$~\msun\ are quite rich in stars, showing therefore a tumultuous stellar accretion history including tidal features such as streams, arcs, and rings superposed on a smoother halo profile. The full movie showing more clearly these features is available in the supplementary material.

\begin{figure*}
\centering
\includegraphics[width=0.49\textwidth]{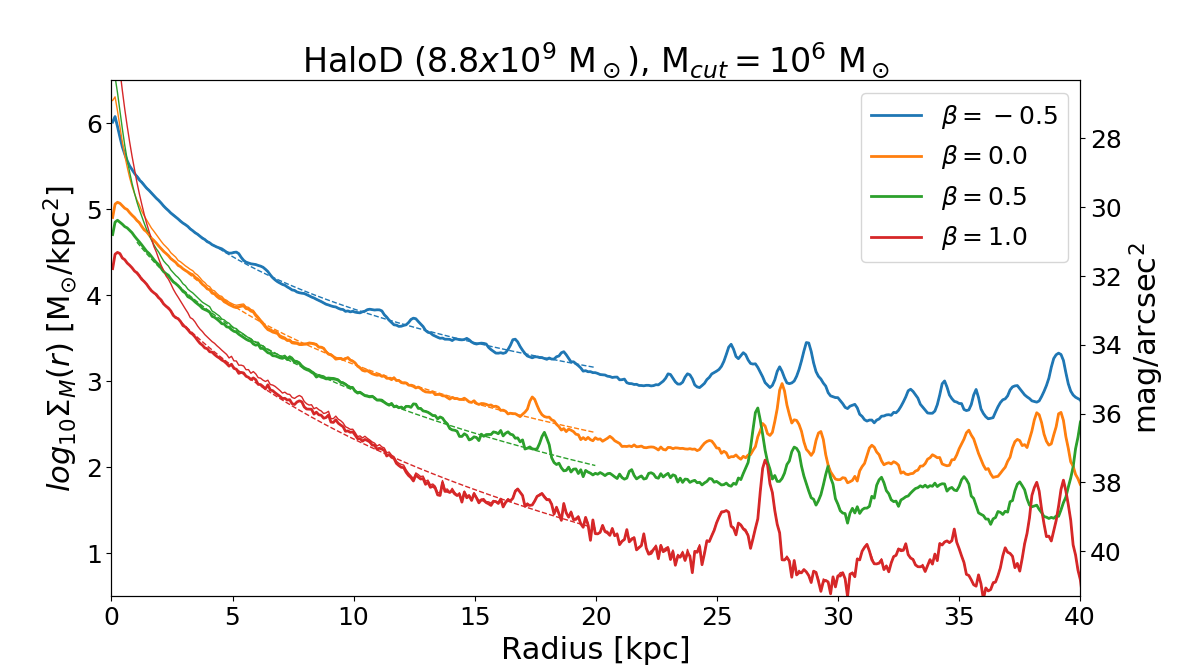}
\includegraphics[width=0.49\textwidth]{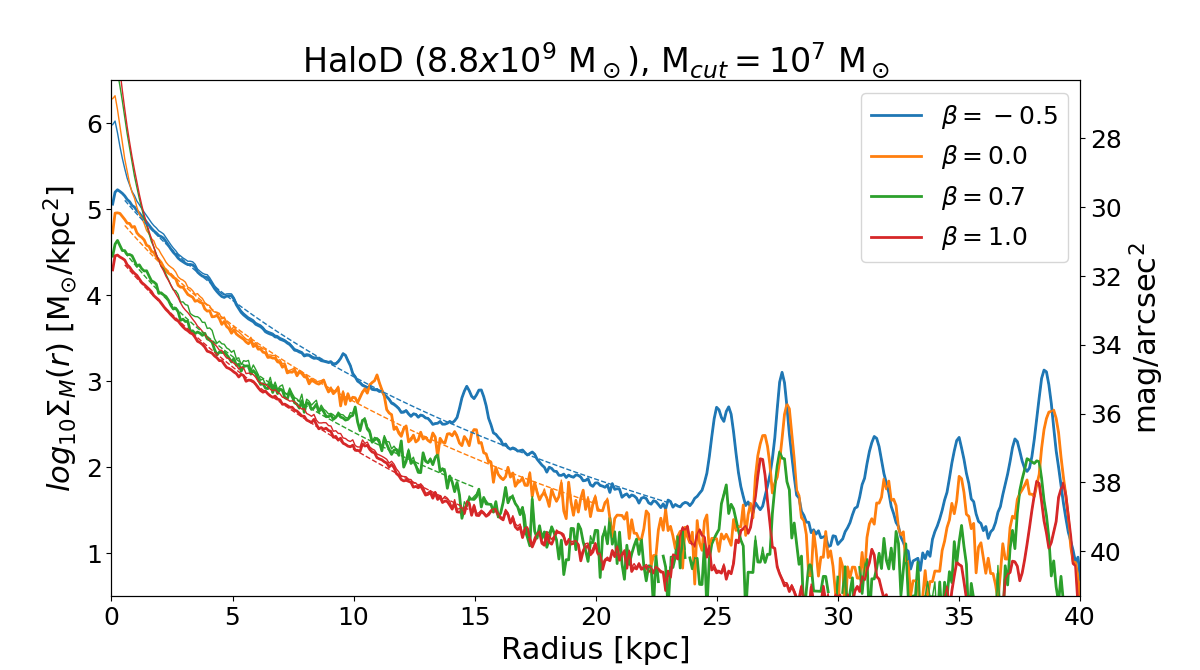}\\
\includegraphics[width=0.49\textwidth]{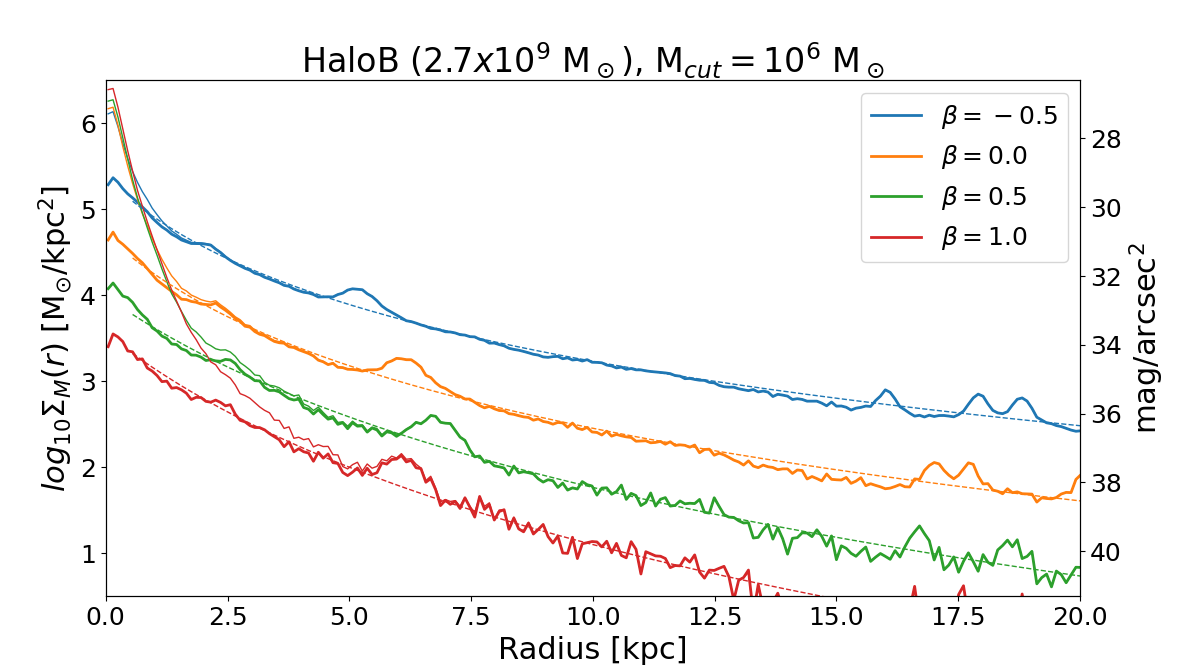}
\includegraphics[width=0.49\textwidth]{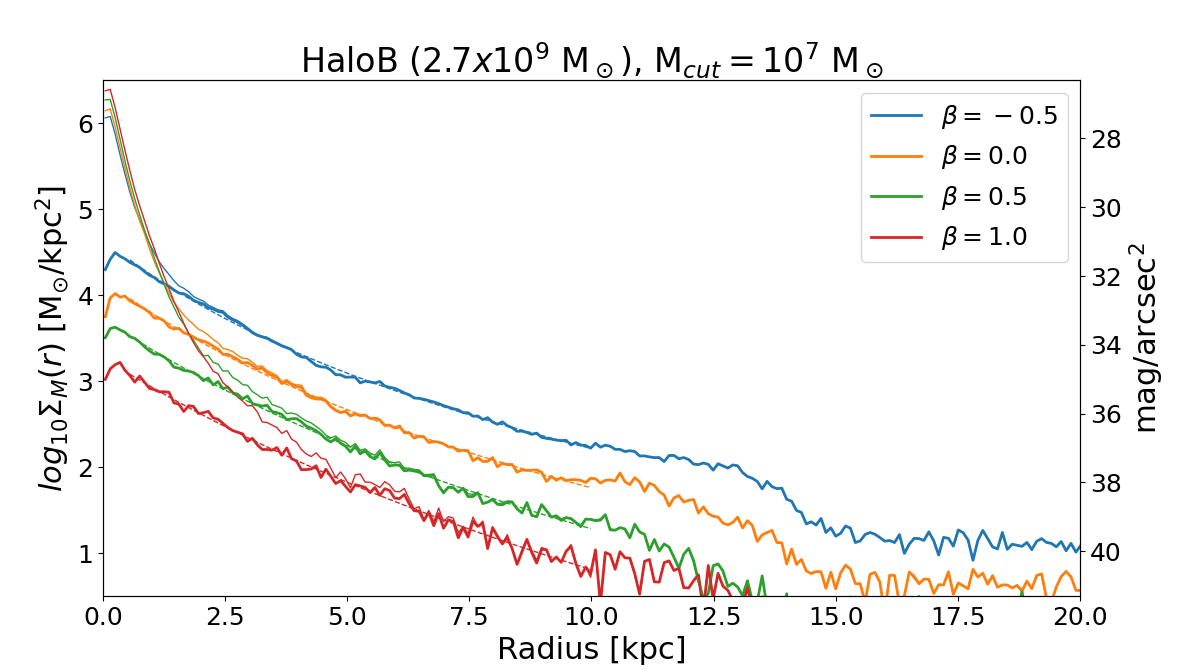}\\
\includegraphics[width=0.49\textwidth]{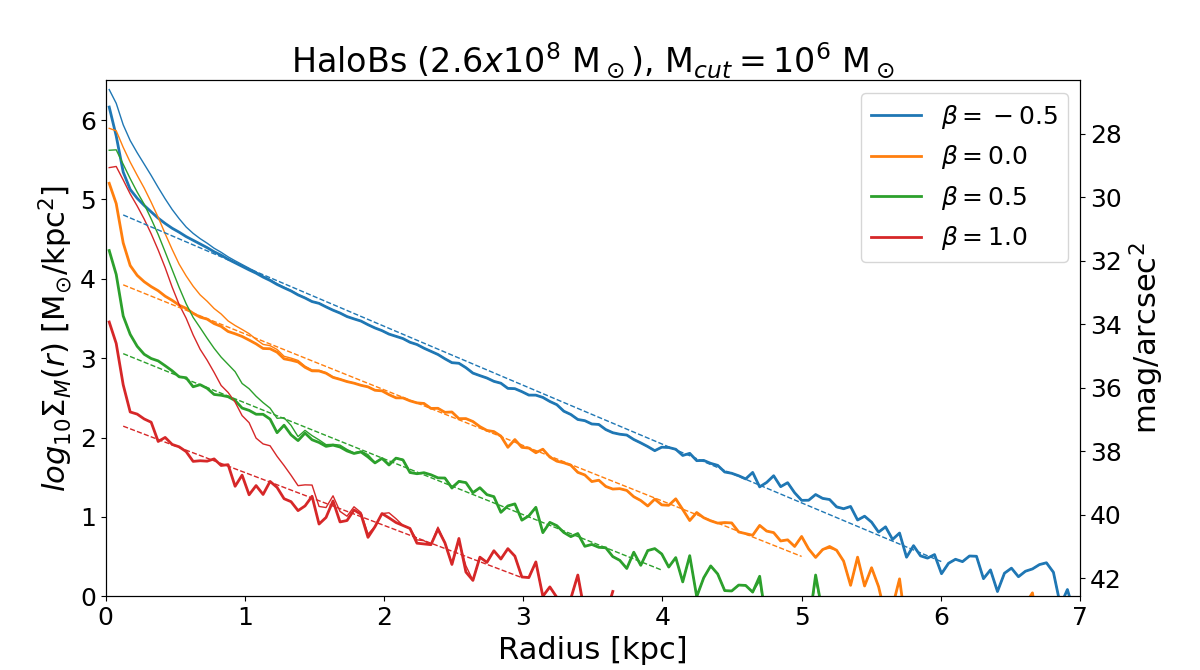}
\includegraphics[width=0.49\textwidth]{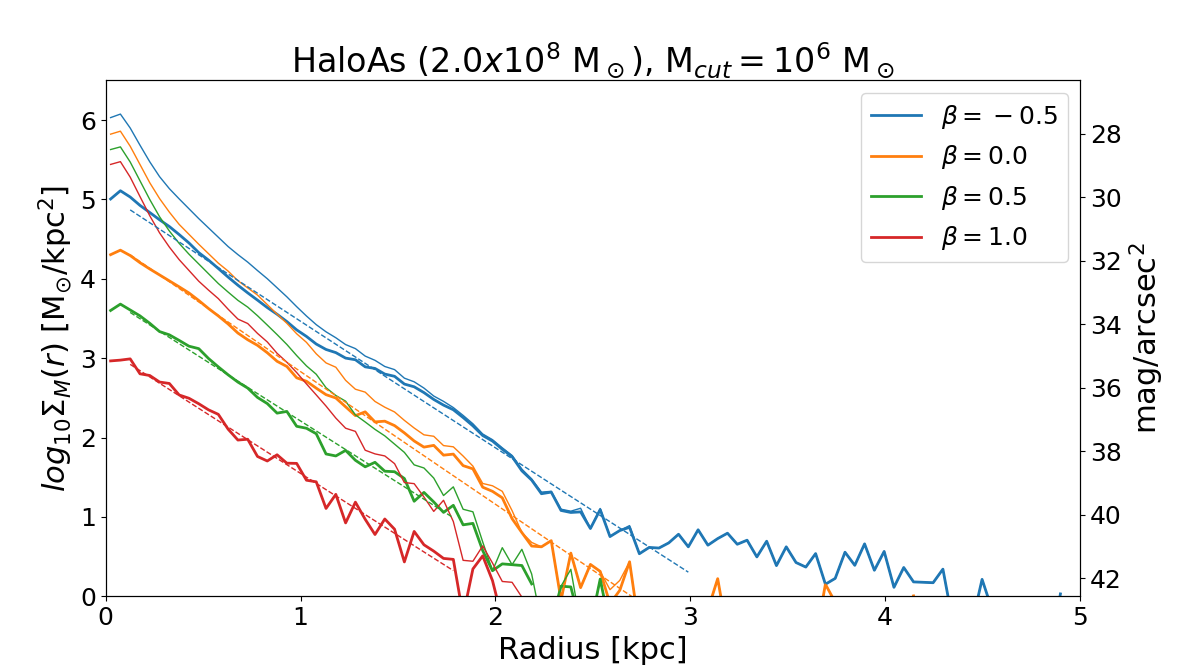}
\caption{Each panel shows the surface density profile of the stars at redshift $z=0$, $\Sigma_M(R)$, as a function of projected radius for different values of $\beta$ (see the legend). The thin solid lines show the azimuthally averaged surface density profiles for the simulated haloes including all stars. The thick solid lines show the same quantity but for the stellar halo stars only (\ie, after removing the stars belonging to central galaxy formed at $z \ge 7$). The thin dotted lines show a fit to the simulation profiles with a generalized exponential function (see text). Each panel, as indicated in its title, refers to haloes with different dark matter masses or assuming different cutoff mass ($M_{cut}$) for the luminous galaxies at $z=7$. Halo As does not have a detectable ghostly stellar halo as the stars from the central galaxy dominate the surface density at all radii.}
\label{fig:profiles}
\end{figure*}

Fig.~\ref{fig:profiles} shows the projected surface density profiles of stars [$\msun$ kpc$^{-2}$] at $z=0$ as a function of distance from the galaxy center for the main set of simulated stellar haloes in Table~\ref{tab:sims}. The solid lines in each panel show the spherically averaged surface density profiles of the stars for a fixed halo mass and $M_{cut}$, changing the values of $\beta$, as shown in the legend. The different panels refer to a different halo mass and cutoff mass as follows: the top panels refer to the most massive halo in our sample (Halo~D) with $M_{cut}=10^6$~\msun\ (top left) and $M_{cut}=10^7$~\msun\ (top right); the middle panels refer to the medium mass halo (Halo~B) with $M_{cut}=10^6$~\msun\ (middle left) and $M_{cut}=10^7$~\msun\ (middle right); the bottom panels refer to the smallest mass haloes (Halo~Bs, bottom left and Halo~As, bottom right), both with $M_{cut}=10^6$~\msun. We show the average of the projected surface densities in the x-, y- and z- directions and we do not remove the surviving satellites from the profile (producing the peaks in the smoother profile especially at large galactocentric distances). The thick solid lines show the profiles excluding the stars in the main progenitor halo\footnote{For Halo~D, in which the central halo merges at high-z with a comparable mass halo, we exclude the the two most massive haloes from the profile.}, while the thin solid lines show the profiles including all the stars. The thin dotted lines show fits to each halo profile (excluding the central most massive halo profile) using the fitting function:
\begin{equation}
\Sigma(R)=\Sigma_0 \left(1+ \frac{1}{\alpha}\frac{R}{R_0} \right)^{-\alpha}.
\label{eq:fits}
\end{equation}
This form of the fitting function is convenient because it converges to an exponential profile [\ie, $\Sigma(R)=\Sigma_0 \exp{(-R/R_0)}$] in the limit $\alpha \rightarrow \infty$. 

The surface density profile of the most massive progenitor retains the exponential profile set in the initial conditions. Since the main haloes we are considering in this study are sufficiently massive to continue accreting gas and forming stars after reionization, we expect that the galaxy in the most massive halo progenitor (\ie, the central galaxy) will grow its stellar mass over time, producing an exponential profile with total mass and scale radius that is larger than its initial value at $z=7$. For this reason we decompose the simulated stellar profiles into two components: the exponential central galaxy profile and the extended stellar halo formed by the debris of all the other satellites. When fitting the observed stellar profiles in isolated dwarfs using our models, we fit simultaneously the exponential profile of the central galaxy and the stellar halo using the model described in \S~\ref{ssec:model}, which excludes the central galaxy contribution. 

Table~\ref{tab:fits} shows the fitting parameters $\Sigma_0$, $R_0$ and $\alpha$ in equation~(\ref{eq:fits}) for the simulations in Fig.~\ref{fig:profiles}. The large values of $\alpha$ in the smallest mass haloes (Halo~Bs and Halo~A), indicate that their stellar haloes have surface density profiles that are very well described by an exponential profile, similarly to the profiles of the central galaxies but with larger scale radii. We will show later that the stellar halo profiles in more massive haloes (Halo~D and Halo~B) deviate from single exponential profiles, but can be understood as the sum of several exponential profiles with different scale radii.

\begin{figure*}
\centering
\includegraphics[width=0.49\textwidth]{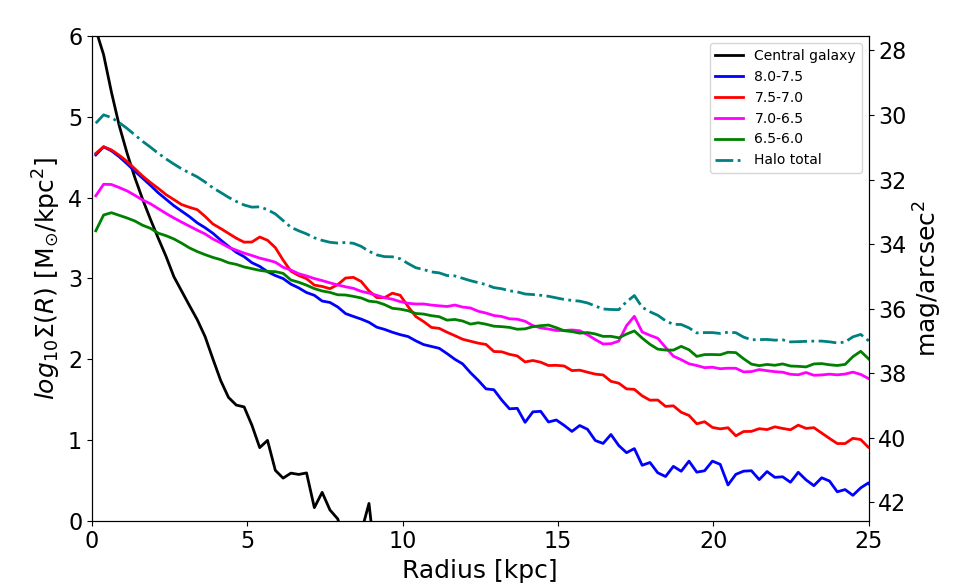}
\includegraphics[width=0.49\textwidth]{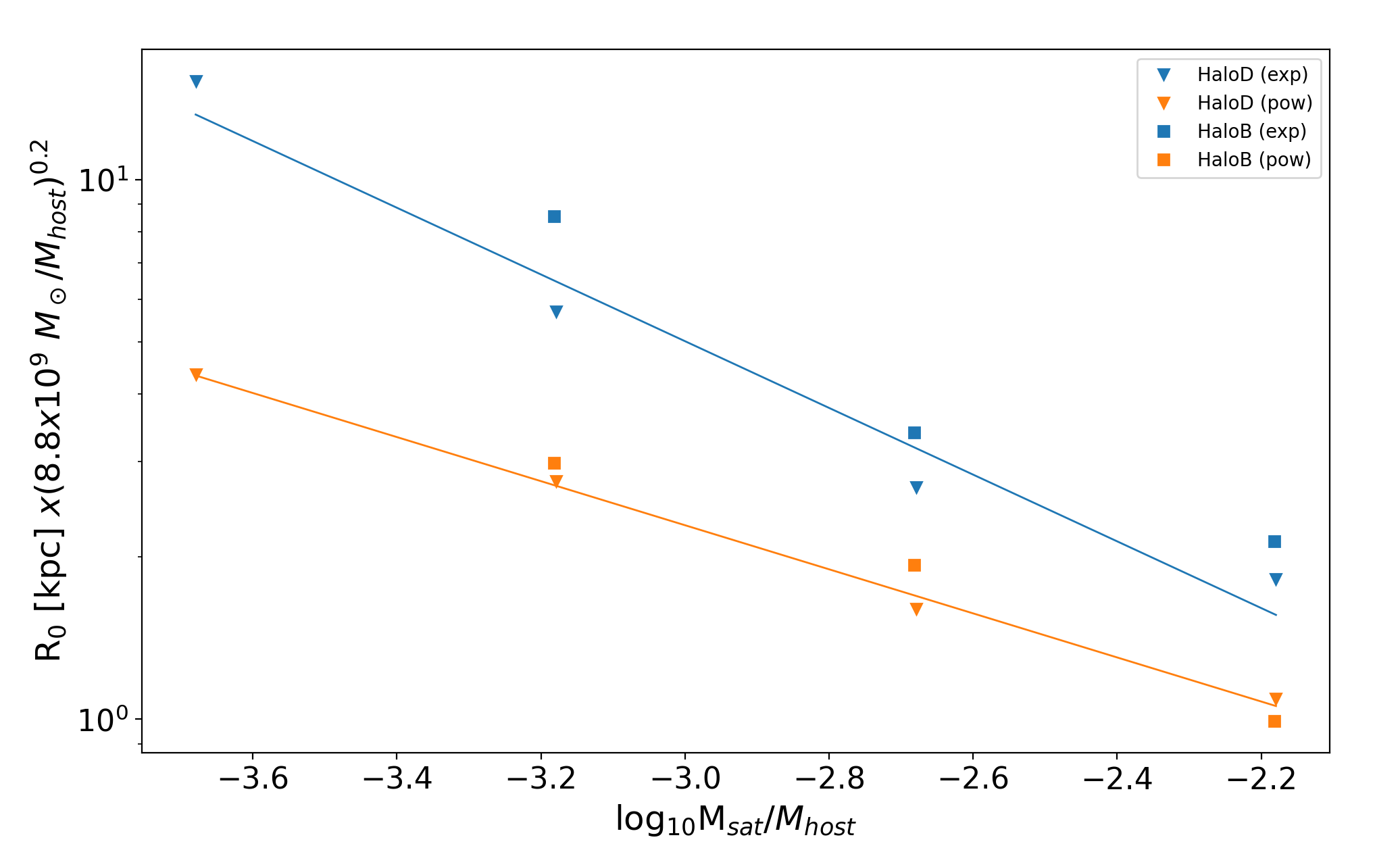}
\caption{{\it (Left)}. Decomposition of the surface density profile of the stars in Halo~D (for $\beta=0$ and $M_{cut}=10^6$~\msun), according to the $z=7$ dark matter mass of the satellites they initially belonged to. The black solid line shows the profile of stars in the central galaxy (the two most massive accreted haloes) and the dotted line shows the stellar halo. The stellar halo is made of several components, each described by a nearly exponential profile with different scale radii $R_0$ shown by the colored solid lines (see legend). The profiles of stars accreted from satellites with dark matter masses $3 \times 10^7<M<10^8$~\msun\ are shown with a blue solid line, $10^7<M<3 \times 10^7$~\msun\ with a red line, $3 \times 10^6<M<10^7$~\msun\  with a purple line, and $10^6<M<3 \times 10^6$~\msun\ with a green line. {\it (Right).} The scale radius $R_0$ of the profile for each of the halo components (as in the left panel) is shown as a function of the dark matter mass log bin of the accreted satellite (normalized to the host halo mass). The triangles refer to Halo D and the squares to Halo B, while the color coding refer to fits of the fundamental profile with an exponential (blue) and with a generalized power-law profile with $\alpha=5$ (orange). The lines show power law fits to the points for Halo~D.
}\label{fig:bins}
\end{figure*}

The dependence of the profile shape on $\beta$ and $M_{cut}$ observed in Fig.~\ref{fig:profiles}, suggests that stars stripped from small mass haloes are preferentially deposited in the outer parts of the profile, and vice versa.
\begin{table}
	\caption{Table of fitting parameters. If the best fit parameter for $\alpha$ is $>500$ we adopt $\alpha=500$.}
	\label{tab:fits}
	\begin{tabular}{lcc|ccc} 
		\hline
		Halo ID & $\log{\left(\frac{M_{cut}}{M_{\odot}}\right)}$ & $\beta$ & $\log{\left(\frac{\Sigma_0}{\rm M_\odot kpc^{-2}}\right)}$ & $\frac{R_0}{\rm {kpc}}$ & $\alpha$\\
		&  &  &  &  & \\
	    \hline
	    
	     &  & -0.5& 5.07 & 0.27 & 500\\
		Halo As & 6 & 0.0 & 4.51 & 0.26 & 500\\
		 &  & 0.5 & 3.77 & 0.28 & 500\\
		 &  & 1.0 & 3.12 & 0.27 & 500\\
		\hline
	  
		 &  & -0.5& 4.90 & 0.58 & 500\\
		Halo Bs & 6 & 0.0 & 4.01 & 0.61 & 500\\
		 &  & 0.5 & 3.15 & 0.61 & 500\\
		 &  & 1.0 & 2.23 & 0.65 & 500\\
		\hline

		  &  & -0.5& 5.37 & 0.77 & 2.88 \\
		Halo B  & 6 & 0.0 & 4.70 & 0.81 & 3.35 \\
		  &  & 0.5 & 3.99 & 1.07 & 4.63 \\
		  &  & 1.0 & 3.56 & 0.95 & 5.02 \\
		\hline
		
		 &  & -0.5& 5.88 & 0.74 & 2.58 \\
		Halo D & 6 & 0.0 & 5.33 & 0.83 & 3.14 \\
		 &  & 0.3 & 5.04 & 0.95 & 3.62 \\
		 &  & 1.0 & 4.71 & 0.93 & 4.48 \\
		\hline
		
		  &  & -0.5& 4.65 & 0.93 & 4.74 \\
		Halo B  & 7 & 0.0 & 4.19 & 1.01 & 5.34 \\
		  &  & 0.5 & 3.73 & 1.12 & 6.56 \\
		  &  & 1.0 & 3.28 & 1.20 & 7.89 \\
		\hline
		
		 &  & -0.5 & 5.29 & 1.23 & 6.06 \\
        Halo D & 7 & 0.0  & 5.00 & 1.23 & 6.06 \\
         &  & 0.7  & 4.70 & 1.12 & 5.46 \\
         &  & 1.0  & 4.54 & 1.27 & 7.38 \\
		\hline
	\end{tabular}
\end{table}
This is more clearly demonstrated in Fig.~\ref{fig:bins} (left), showing the stellar surface density profile of Halo~D (with $\beta=0$ and $M_{cut}=10^6$~\msun), for a subset of stars coming from satellite haloes in the mass range $\log M$ to $\log M + \Delta \log M$, for different values of $\log M$ as shown in the legend. The black solid line shows the profile of stars in the central galaxy (the two most massive haloes) and the dotted line shows the stellar halo. The profiles of stars accreted from satellites with dark matter masses $3 \times 10^7$~\msun\ $< M <10^8$~\msun\ are shown with a blue solid line, $10^7$~\msun\ $< M < 3 \times 10^7$~\msun\ with a red solid line, $3 \times 10^6$~\msun\ $< M < 10^7$~\msun\ with a purple solid line, and $10^6$~\msun\ $< M < 3 \times 10^6$~\msun\ with a green solid line. 
The figure clearly shows that the outer parts of the stellar halo are contributed from stars stripped from smallest mass satellites: each mass bin produces a profile that can be approximated by an exponential with scale radius $R_0$ that is larger for smaller dark matter halo masses of the accreting satellites.
The right panel in Fig.~\ref{fig:bins} shows the scale radius $R_0$ as a function of the dark matter mass log bin of the accreted satellites. The symbols show $R_0 \times (8.8\times 10^9~M_\odot/M_{host})^{0.2}$ as a function of $M_{sat}/M_{host}$ for Halo~D (triangles) and Halo B (squares) for simulations with $\beta=0$ and $M_{cut}=10^6$~\msun. The lines show the power law fit to the points for Halo D and the color of the symbols and lines refer to $R_0$ obtained fitting the profiles with an exponential profile (blue) and with a generalized power-law profile [equation~(\ref{eq:fits})] with $\alpha=5$ (orange). Comparing the points for Halo D and Halo B we observe that the scale radius $R_0$ depends on the host halo dark matter mass as a power-law with slope slightly shallower than the expected $1/3$ slope: $R_0 \propto M_{host}^{0.2}$.

Fitting the fundamental bin profiles with an exponential function we found the following power-law fits to the scale radius $R_0$ as a function of the satellite mass $M_{sat}$ and the host halo mass $M_{host}$:
\begin{equation}
R_0^{\rm (exp)}=0.07~{\rm kpc}\left(\frac{M_{host}}{8.8\times 10^9~M_\odot}\right)^{0.2}\left(\frac{M_{sat}}{M_{host}}\right)^{-0.62}.
\end{equation}
Fitting the fundamental bin profiles with a generalized power-law profile with $\alpha=5$ we find:
\begin{equation}
R_0^{\rm (pow)}=0.137~{\rm kpc}\left(\frac{M_{host}}{8.8\times 10^9~M_\odot}\right)^{0.2}\left(\frac{M_{sat}}{M_{host}}\right)^{-0.4}.
\end{equation}
This quantitative fitting of $R_0$ as a function of the mass of the satellites, $M_{sat}$, will be the foundation to build our halo model in \S~\ref{ssec:model}.
We speculate that the physical reason for the dependence of the scale radius on the ratio of the satellite to host halo mass is due to the interplay of dynamical friction and tidal destruction. Mergers with mass ratios closer to unity are rapidly spiraling in (in few crossing times), depositing most of the dark matter and stars in the inner parts of the haloes. Vice versa, small mass satellites (smaller mass ratio mergers) orbit the galaxies for several crossing times before being completely tidally stripped, depositing the stars further out.

Finally, the dependence of the profiles on $\epsilon_0$ in equation~(\ref{eq:fstar}) is not shown in the panels because it is trivial: the whole surface density profile is simply proportional to $\epsilon_0$.

\subsubsection{Dependence on the compactness of satellites and Cosmic Variance}\label{ssec:reff}

\begin{figure*}
\centering
\includegraphics[width=0.48\textwidth]{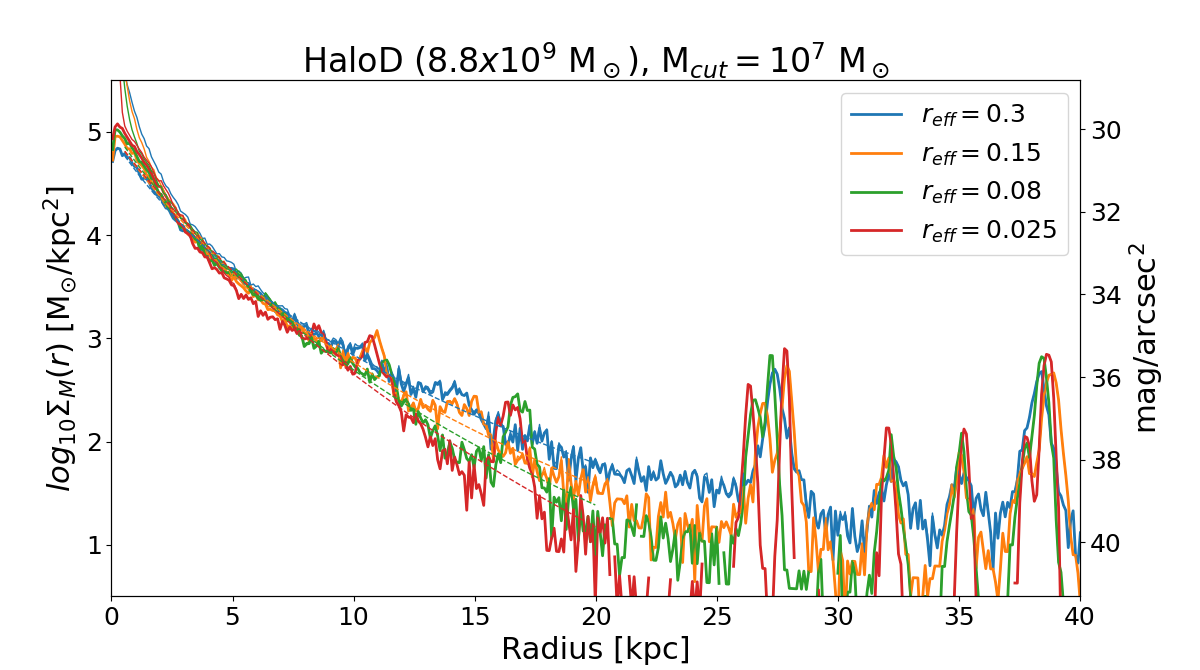}
\includegraphics[width=0.48\textwidth]{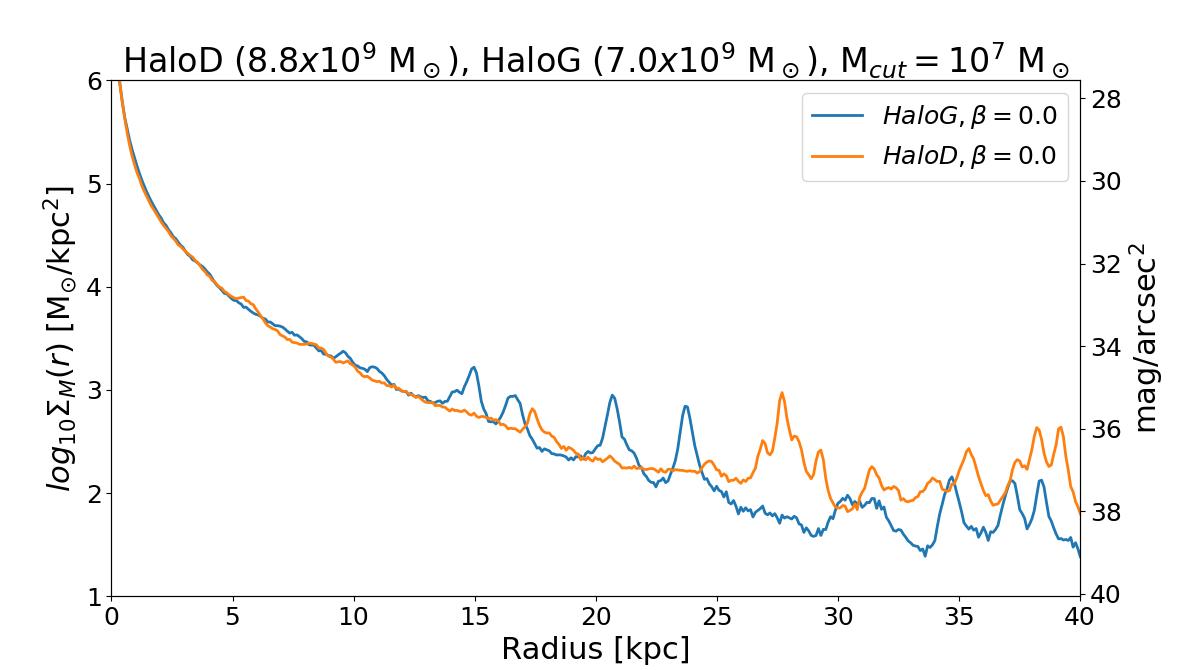}
\caption{{\it (Left)}. Same as Fig.~\ref{fig:profiles} but for Halo~D with $\beta=0$ and $M_{cut}=10^7$~\msun\ and different values of r$_{eff}$ as in the legend. This comparison shows that at small to intermediate radii (\ie, at radii with available observations) the assumed compactness of merging galaxies does not have a strong effect on the surface density profile. The outer parts of the halo ($R>20$~kpc), however, with surface brightness much lower than what accessible to observations, differ depending on $r_{eff}$. {\it (Right)}. Comparison between Halo~D (mass $8.8 \times 10^9$~\msun) and Halo~G (mass $7.0 \times 10^9$~\msun) with $\beta=0$ and $M_{cut}=10^6$~\msun\ shows that for haloes of this mass the effect of cosmic variance appears small, even thought more simulations are needed to conclude that it is negligible.}
\label{fig:profiles1}
\end{figure*}

The left panel in Fig.~\ref{fig:profiles1} shows the surface density profile of stars in Halo~D for simulations with $\beta=0$, $M_{cut}=10^7$~M$_\odot$, and different values of the effective radius of the stars in the initial conditions at $z=7$. As explained in \S~\ref{sec:sim}, the stars in each dark matter halo at $z=7$ have an exponential profile with scale radius $r_{eff}$ proportional to the virial radius $r_{vir}$ as determined by the parameter $\eta$. We adopted a fiducial value $\eta=0.15$ at $z=7$. The different lines in the left panel of Fig.~\ref{fig:profiles1} refer to $\eta=0.3, 0.15, 0.08, 0.025$, as shown in the legend. This comparison shows that the compactness of merging galaxies does not have a strong effect on the surface density profile at small to intermediate radii (\ie, at radii at which observations are either available or feasible in the near future using stellar counts). Instead, the outer parts of the stellar halo at $R>20$~kpc, where the surface brightness is much lower than what is testable with observations, differ depending on $r_{eff}$ in the expected direction according to our qualitative model: if the stars in the accreting satellites are more concentrated (\eg, accreted globular clusters) then less stars are deposited by tidal stripping in the outer stellar halo. Vice versa, if the stars are more diffuse in the accreted dwarf galaxy (\eg, accreted UF dwarf galaxies), more stars are stripped in the outer regions of the stellar halo and the outer profile has higher surface density. We also note that the number of surviving satellites depends on the compactness of the stellar component in the satellites, as expected.

The right panel in Fig.~\ref{fig:profiles1} 
shows the comparison between two haloes with similar mass at $z=0$ to test cosmic variance and variations of the merger history: Halo~D (mass $8.8 \times 10^9$~\msun) and Halo~G (mass $7.0 \times 10^9$~\msun) assuming $\beta=0$ and $M_{cut}=10^7$~\msun. This result suggests that the stellar halo profiles are nearly indistinguishable in dark matter haloes with masses $>10^9-10^{10}$~\msun. Clearly a larger sample is necessary to conclude that the effect of cosmic variance is negligible in this mass range, but this result already gives an indication of the possible small amplitude of the scatter. However, in our simulations we assumed a monotonic mass to light ratio of galaxies at $z=7$ with no scatter and 100\% occupancy (no dark haloes). This assumption is clearly simplistic, hence we expect the variance of properties in our test to be underestimated.

For Halo~As and Halo~Bs that have a mass $\sim 2 \times 10^8$~\msun\ and $<10$ luminous satellites the variance is very large (see bottom panels in Fig.~\ref{fig:profiles}). Halo~As has 2 satellites and does not have a significant stellar halo, while Halo~Bs has 6 satellites and the stellar halo can be detected if $f_* > 10^{-3}$ in equation~(\ref{eq:fstar}).

\subsection{Modelling the Halo Profile}\label{ssec:model}

In this section we present a model of the surface density profiles of stellar haloes in the simulations following the findings, discussed in \S~\ref{ssec:profiles}, that the profile can be understood as the sum of nearly exponential profiles (or generalized power-law profiles) with scale radii determined by the mass of the satellite building it. Namely, with small mass haloes depositing most of their stars further out in the stellar halo, and vice versa. We therefore write the surface density profile in the model as sum (or an integral in the continuum limit) of the surface density profiles contributed by each log mass bin of satellites building the halo:
\begin{equation}
\Sigma_M(R,M_{host}) = \int_{M_{cut}}^{M_{max}} \Sigma_{bins}(R,M_{host},M_{sat}) d \ln M_{sat},
\label{eq:model}
\end{equation}
where the limits of integration are $M_{cut}$ for the lower limit and the maximum mass of the accreted satellite excluding the host halo, $M_{max}$, for the upper limit.
For the fundamental shape of the surface density profiles in each mass bin, we can use an exponential profile:
\begin{equation}
\Sigma^{\rm (exp)}_{bins}(R,R_0)=\frac{dM_*}{d\ln M_{sat}}\frac{1}{2\pi R_0^2}\exp{\left(-\frac{R}{R_0}\right)}
\label{eq:exprof}
\end{equation}
or a generalized power-law profile with exponent $\alpha$:
\begin{equation}
\Sigma^{\rm (pow)}_{bins}(R,R_0)=\frac{dM_*}{d \ln M_{sat}}\frac{{\cal A}}{2\pi(\alpha R_0)^2}\left(1+\frac{1}{\alpha}\frac{R}{R_0}\right)^{-\alpha},
\end{equation}
that in the limit $\alpha \rightarrow \infty$ converges to equation~(\ref{eq:exprof}). The normalization constant is ${\cal A} \equiv [(\alpha-2)^{-1}-(\alpha-1)^{-1}]^{-1}$. 
Here, $dM_*/d \log M_{sat}$ is the mass in stars contributed by satellites with masses between $\log M_{sat}$ and $\log M_{sat} + d \log M_{sat}$: 
\begin{equation}
\frac{dM_*}{d \ln M_{sat}} (M_{host},M_{sat})=f_* M_{sat} \frac{dN_{sat}}{d\ln M_{sat}},
\end{equation}
where $dN_{sat}/d\log M_{sat}$ is the halo mass function of the merging satellites.

\begin{figure}
\centering
\includegraphics[width=0.48\textwidth]{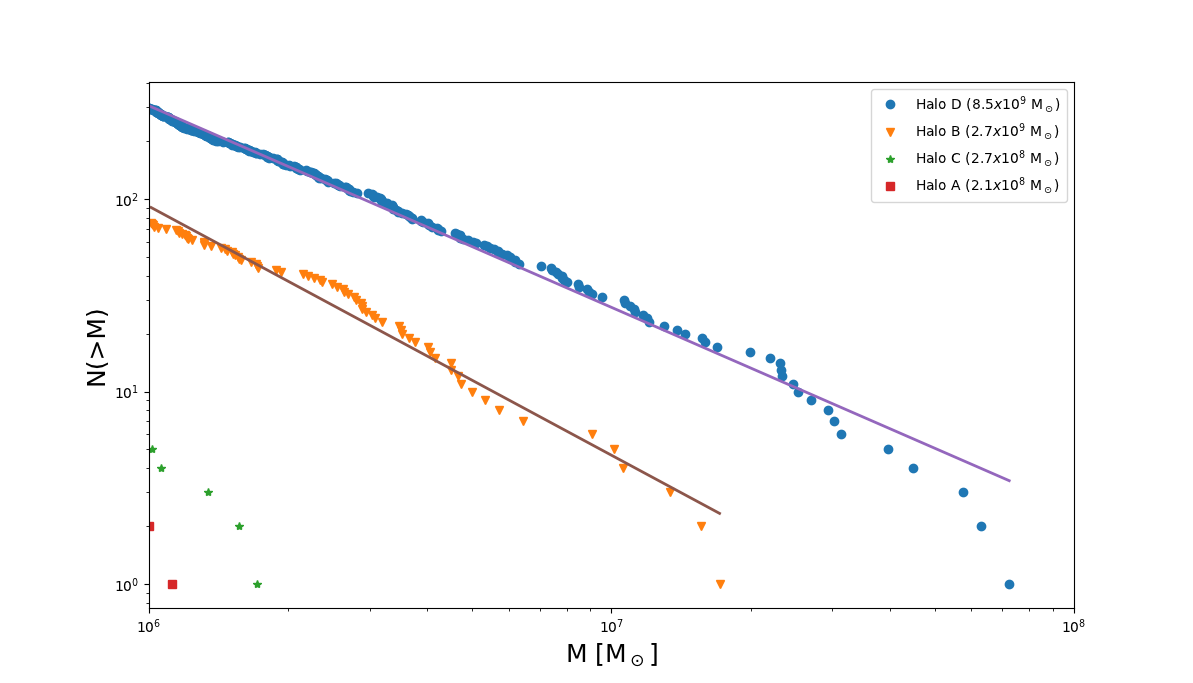}
\caption{The points show the cumulative dark matter mass function ($N(>M)$) of progenitor masses at $z=7$, merging to form the $z=0$ Halo~D, Halo~B, Halo~Bs and Halo~As (see the legend). The solid lines show power-law fits to the mass functions for the two most massive haloes. Halo~Bs has only 6 satellites with $z=7$ masses $>10^6$~M$_\odot$, while Halo~As has two.}\label{fig:massfunc}
\end{figure}

Fig.~\ref{fig:massfunc} shows cumulative dark matter mass function of progenitors masses at $z=7$ merging to form the $z=0$ Halo~D, Halo~B, Halo~Bs and Halo~As (see legend). The solid lines show power-law fits to the mass function for the two most massive haloes: Halo~D, $N_{sat}(>M)=309 (M/10^6~M_{\odot})^{-1.05}$, and Halo~B $N_{sat}(>M)=91  (M/10^6~M_{\odot})^{-1.29}$. The number of satellites for Halo~Bs and Halo~As are 6 and 2, respectively.
In the analytic model we adopt a single power law form for all the halo masses that is a good fit to Halo~D and Halo~B:
$N_{sat}(>M)=N_{tot}(M/10^6~M_{\odot})^{-\alpha}$, with $\alpha=1.2$ and $N_{tot}=76 (M_{host}/2.7\times 10^9 M_\odot)$, reproducing the total number of satellites in Halo~D.
The relationship,
\begin{equation}
\frac{dN_{sat}}{d \ln M_{sat}}=\alpha N_{sat}(>M) \approx 1.2 N_{tot}\left(\frac{M_{sat}}{10^6~M_\odot}\right)^{-1.2},
\end{equation}
closes the system of equations in our model.
Next, we test whether the model in equation~(\ref{eq:model}), constructed dissecting the simulation results, actually reproduces the stellar haloes in all the simulations in Table~\ref{tab:sims}.

\begin{figure*}
	\centering
	\includegraphics[trim=0 0 0 0, clip, width=2.0\columnwidth]{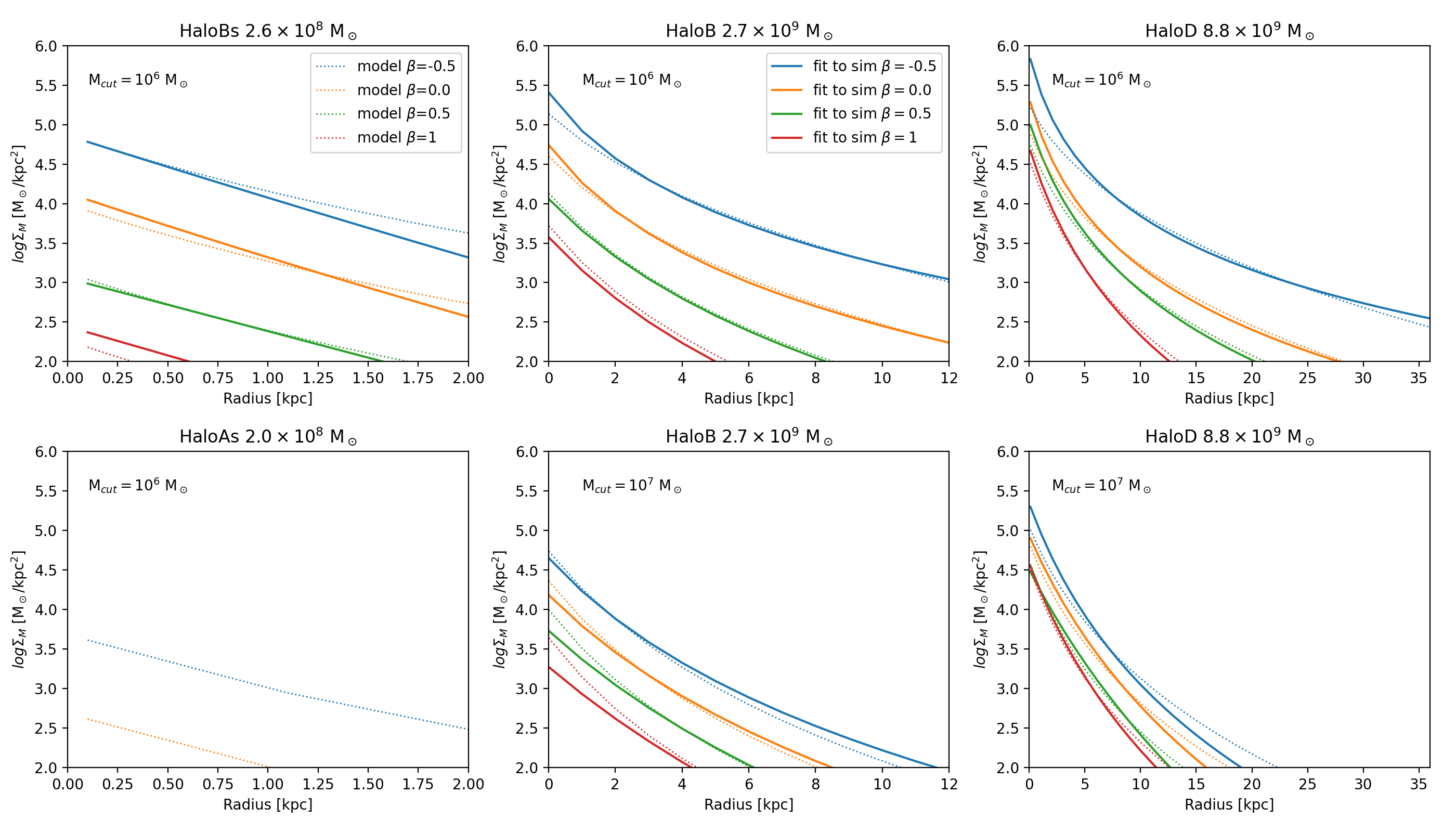}
    \caption{Ghostly stellar halo model (dotted lines) compared to fits to the simulation results (solid lines) for different values of $\beta$ as shown by the legend. In the top row, models and fits to the simulations assume $M_{cut}=10^6$~M$_\odot$ and from left to right, each column shows haloes of increasing mass at $z=0$: Halo~Bs of mass $2.6\times 10^8$~\msun\ (top-left), Halo~B of mass $2.7 \times 10^9$~\msun\ (top-center), and Halo~D of mass $8.8 \times 10^9$~\msun\ (top-right). The panels in the bottom row refer to Halo~As ($2.0\times 10^8$~\msun, bottom-left) with $M_{cut}=10^6$~\msun\, Halo~B (bottom-middle) and Halo~D (bottom-right) both assuming $M_{cut}=10^7$~\msun. Note that for the smallest mass halo (Halo~As) an extended stellar halo does not exist, and it exists only for $M_{cut}=10^6$~M$_\odot$ in Halo~Bs.}
    \label{fig:model}
\end{figure*}

In Fig.~\ref{fig:model}, we show the ghostly halo model (dotted lines) compared to fits (see equation~[\ref{eq:fits}] and Table~\ref{tab:fits}) to the simulation results (solid lines) for different values of $\beta$ as shown in the legend. From left to right, each column refer to haloes with increasing mass at $z=0$: Halo~Bs of mass $2.6\times 10^8$~\msun\ (top left panel), Halo~As of mass $2.0 \times 10^8$~\msun\ (bottom left panel), Halo~B of mass $2.7 \times 10^9$~\msun\ (center panels), Halo~D of mass $8.8 \times 10^9$~\msun\ (right panels). The center and right panels in the top and bottom rows refer to haloes with $M_{cut}=10^6$~\msun\ and $M_{cut}=10^7$~\msun, respectively, while the model/simulations in left panels have $M_{cut}=10^6$~\msun. Here we used the generalized profile with $\alpha=5$ for the fundamental surface density profiles for each mass bin in equation~(\ref{eq:model}), but assuming exponential profiles produces only slightly worse agreement at small radii between the model and the fits to the simulated halo profiles.
This model is empirical but physically motivated and, as illustrated in Fig.~\ref{fig:model}, it reproduces the simulation results accurately.

In the next section, armed with this model for the surface density profile of stellar haloes in dwarf galaxies, we apply it to fit observed data on the stellar surface density profiles of six Local Group galaxies, and consequently derive the model's free parameters. 

Finally, for the sake of comparing our results to published studies based on halo-matching methods, we are interested in the relationship between the dark matter halo masses at $z=7$ of progenitor satellites and the maximum mass such satellites can reach before starting to lose mass while merging with the main host halo. 
Since the satellites in our simulations form all their stars before the epoch of reionization (see Fig.~\ref{fig:mergerhist}), their stellar masses remain constant but their dark matter masses generally increase from the value at $z=7$. Hence, $f_*(z)\equiv M_*(z=7)/M_{dm}(z)$ can decrease with respect to the initial value at $z=7$.  

\begin{figure*}
\centering
\includegraphics[width=\textwidth]{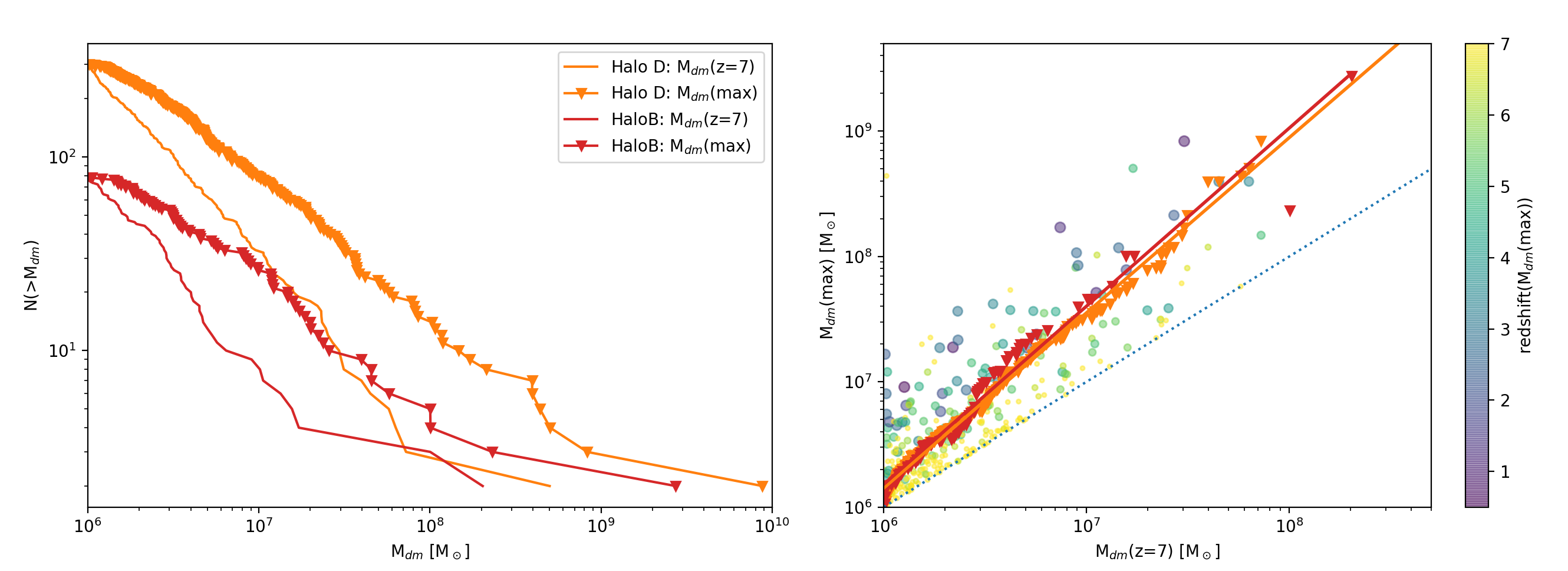}
\caption{{\it (Left).} Cumulative number of satellites in Halo~D (orange lines) and Halo~B (red lines) at $z-0$ as a function of their dark matter mass at $z=7$, $M_{dm}(z=7)$ (solid lines) and their maximum dark matter mass, $M_{dm}(max)$ (triangles). {\it (Right).} The circles show the maximum dark matter mass of satellites, $M_{dm}(max)$ as a function of $M_{dm}(z=7)$ for each subhalo in Halo~D and Halo~B. The triangles show the relationship between $M_{dm}(max)$ and $M_{dm}(z=7)$ by matching the two cumulative distribution values with the same $N(>M_{dm})$, shown in the left panel. The solid lines show the power-law fits to the data points for the two haloes. The dotted line show $M_{dm}(max)=M_{dm}(z=7)$ for comparison. The color bar on the right refer to the redshift at which $M_{dm}$ reaches its maximum value.}
\label{fig:mass_mass}
\end{figure*}

The left panels in Fig.~\ref{fig:mass_mass} shows the cumulative number of satellites in Halo~D (orange lines) and Halo~B (red lines) at $z=0$ as a function of their dark matter mass at $z=7$ (solid lines) and as a function of their maximum dark matter mass, $M_{dm}(max)$ (triangles). The curves illustrate how the masses of dark matter haloes increase from $z=7$ to the time when they merge and start to lose mass. The right panel in Fig.~\ref{fig:mass_mass} shows the maximum dark matter mass of satellites, $M_{dm}(max)$ as a function of $M_{dm}(z=7)$ for the same haloes as in the left panel, illustrated in two different ways: i) The circles show $M_{dm}(max)$ as a function of $M_{dm}(z=7)$ for each subhalo in Halo~D and Halo~B. The color coding of the circles refers to the redshift at which the halo mass reaches its maximum value, as shown by the colorbar. ii) The orange and red triangles show the relationship between $M_{dm}(max)$ and $M_{dm}(z=7)$ by matching the two cumulative distributions in the left panel with the same $N(>M_{dm})$, for Halo~D and Halo~B, respectively. The solid lines show power-law fits to the triangle data points for the two haloes. For comparison, the dotted line shows $M_{dm}(max)=M_{dm}(z=7)$. 

Since the power law fits to both haloes are the same within the statistical error, we can therefore use the power law fit,
\begin{equation}
M_{dm}(max)=3.55\times 10^7~M_\odot \left(\frac{M_{dm}(z=7)}{10^7~M_\odot}\right)^{1.4},\label{eq:Mmax}
\end{equation}
to convert (statistically) $M_{dm}(z=7)$ into $M_{dm}(max)$.

In summary, since all the satellites haloes have maximum dark matter mass below the threshold for gas accretion and star formation due to reionization of the IGM, we can assume that in this halo mass range $M_*(max)= M_*(z=7)$: all the stars in these dwarfs galaxies have formed before reionization, consistently with them being UF dwarfs, fossils of the first galaxies \citep{BovillR2009}. The relationship between $M_*$ and the maximum dark matter halo mass of satellite haloes will be useful to compare our results in the next section to published works based on sophisticated halo-matching methods \citep[\eg,][]{Behroozi2013, Nadleretal:2020}.

\section{Constraints on the Star Formation Efficiency in the First Galaxies}\label{sec:constraints}

\citet{KangR:2019} have compiled a list of six dwarf galaxies (Leo~T, Leo~A, WLM, IC~1613, IC~10, and NGC~6822) located in the Local Group, but outside the virial radii of the Milky Way and Andromeda, showing robust or tentative evidence of the existence of an extended stellar halo.
In the collected observational data sets, radial density profiles of stars are provided out to several half-light radii from the central galaxy. The stellar haloes are found by star counts, and usually only the red giants (RGs) are sufficiently bright to be detected in deep colour–magnitude diagrams (CMD). Since the data are in terms of number of RG stars selected around the tip of the RG branch, this observed number needs to be converted into a stellar mass at zero-age main sequence (at the time of the galaxy formation) that in our case coincides with redshift $z \sim 7$. We refer to \cite{KangR:2019} for details on the conversion using synthetic CMDs, assuming stellar evolutionary tracks appropriate for dwarf galaxies that formed all their stars before reionization. Here we use the data points for each galaxy profile from \citet{KangR:2019}, including the conversion between number of RG stars and stellar mass at formation (see their Table~3 for stellar halo parameters). 
Estimates of the total dark matter halo masses of the six observed dwarfs are also derived in \citet{KangR:2019} (see their Table~4). For each dwarf galaxy, estimates of the total dark matter halo mass was derived taking the average of the masses obtained using three different independent approaches: i) Estimates found in the literature, when available. But the discrepancy between different authors can be up to one order of magnitude; ii) A method based on the knowledge of the stellar mass of the dwarf galaxy at $z = 0$; iii) A method based on the knowledge of the half-light radius. We refer to the original paper for details.

\begin{figure*}
\centering
\includegraphics[width=\textwidth]{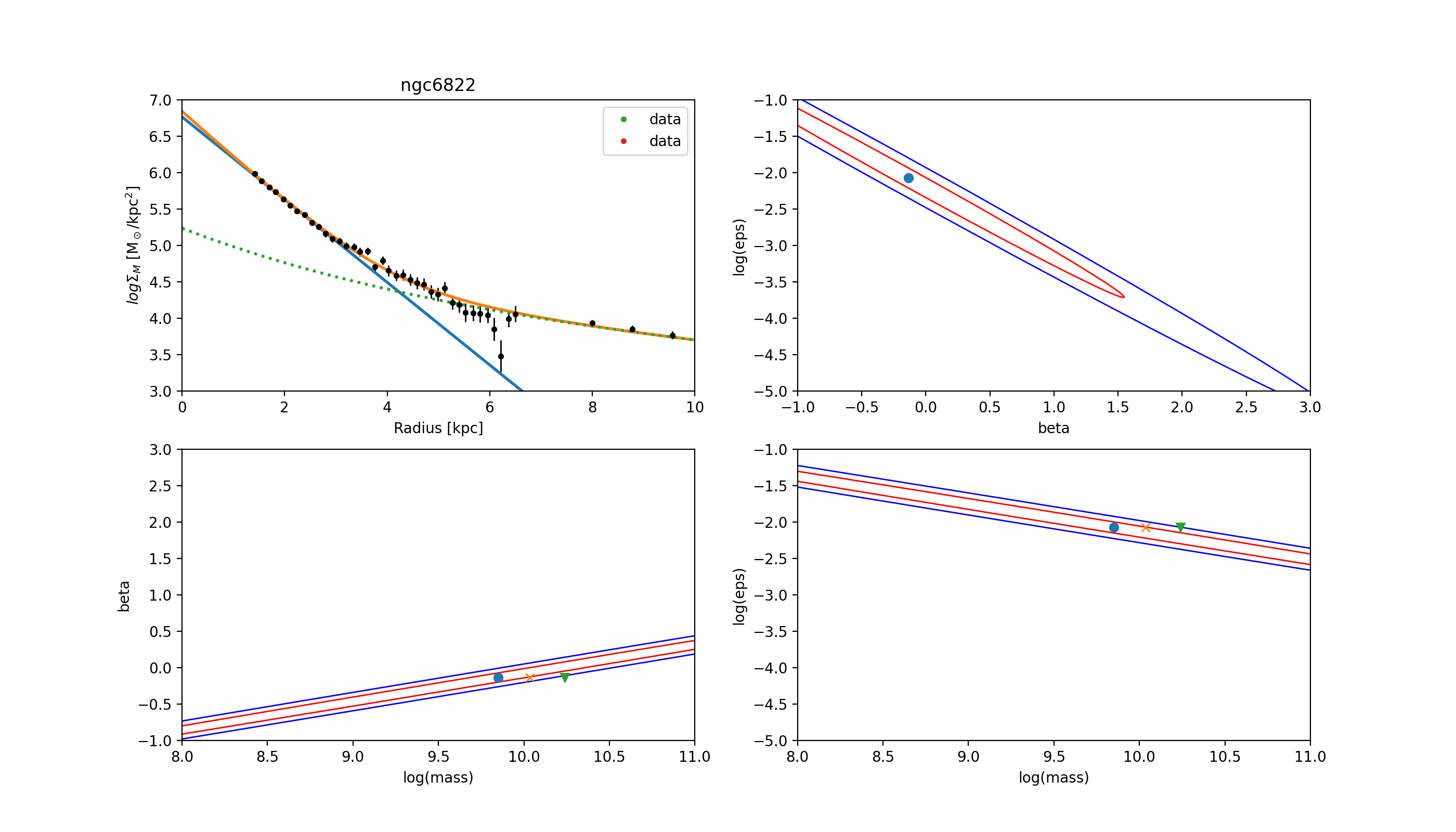}
\caption{Fit of the ghostly halo model presented in this work to the surface density profile data for NGC~6822. {\it (Top left.)} Surface density profile of NGC~6822 (points), best fit exponential profile of the galaxy (solid blue line), best fit ghostly halo model (dotted line), and total galaxy plus halo best fit model of surface density profile (solid orange line). {\it (Top right.)} Confidence contour plots of the fitting parameters $\beta$ and $\epsilon_0$ (red 68\%, and blue 95\% confidence). {\it (Bottom left.)} Confidence contour plots for the fitting parameters  $M_{halo}$ and $\beta$. {\it (Bottom right.)} Confidence contour plots for the fitting parameters $M_{halo}$ and $\epsilon_0$.}
\label{fig:ngc6822}
\end{figure*}

The free parameters in our ghostly halo model are four. Three of them: $\beta$, $\epsilon_0$ and $M_{cut}$, describe the star formation efficiency [see equation~(\ref{eq:fstar})] in primordial galaxies before the epoch of reionization at $z\sim 7$. The other free parameter is the host halo dark matter mass, $M_{halo}$, at $z=0$. 

In addition to the four free parameters describing the ghostly stellar haloes, we need to separate the halo stars from the stars belonging to the central galaxy. To do this we fit the central galaxy surface brightness with an exponential profile: $\Sigma^{gal}=\Sigma_0^{gal}\exp{(-R/R_{exp}^{gal})}$. Hence, we have two more free parameters: $\Sigma_0^{gal}$ and $R_{exp}^{gal}$. Unless otherwise stated, we fit the data points to the halo model and the exponential galaxy profile at the same time, using a maximum likelihood estimator with $4+2=6$ free parameters.

\begin{figure*}
\centering
\includegraphics[width=\textwidth]{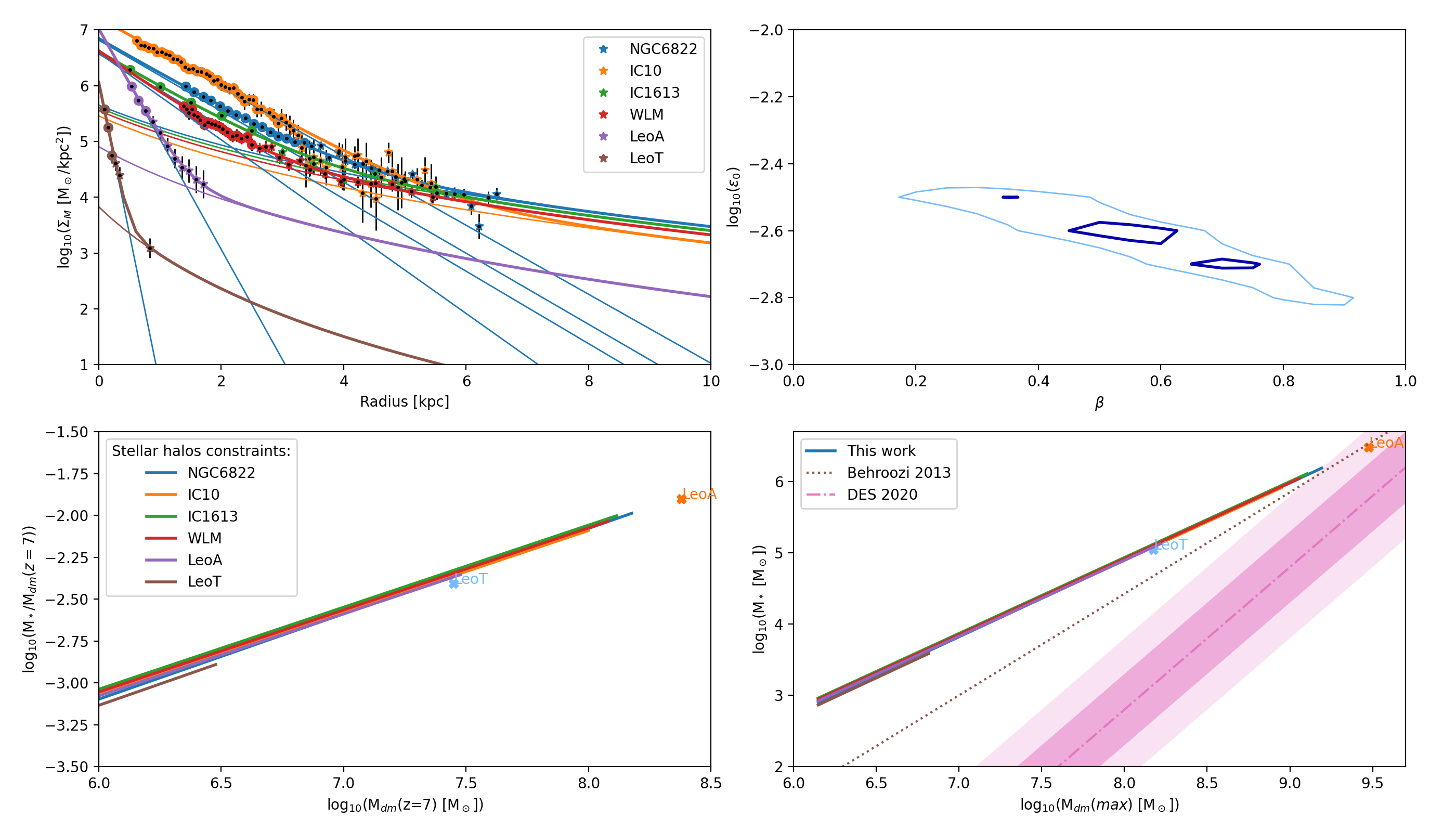}
\caption{{\it (Top left).} Observed surface density profiles of the stars in six isolated dwarf galaxies in the Local group (data points, see legend) and fits to the exponential galaxy profile (blue solid lines) and the ghostly halo model in this work (colored solid lines). {\it (Top right).} $1\sigma$ and $2\sigma$ confidence contour plots for the fitting parameters $\beta$ and $\epsilon_0$, where $f_*=\epsilon_0 (M_{halo}/10^7~\msun)^\beta$. 
{\it (Bottom left).} Stellar mass as a function of the $z=7$ dark matter halo mass. Each solid line refer to the results of this work for a different isolated dwarf galaxy considered in this study (see legend). The symbols show the stellar mass and $z=7$ halo mass of Leo~T and Leo~A. {\it (Bottom right).} Stellar mass as a function of the maximum value of the dark matter mass of halos. The solid lines refer to the results of this work, the dotted line refer to results from \protect{\citet{Behroozi2013}}, 
and the shaded area refer to the DES results \protect{\citep{Nadleretal:2020}}. 
The symbols show the stellar masses and maximum halo masses of Leo~T and Leo~A (see text for more details).}
\label{fig:results}
\end{figure*}

Fig.~\ref{fig:ngc6822} shows, for purely illustrative purposes, the fit of our ghostly halo model presented in this work, to the observed surface density profile data for NGC~6822, with the following caveat: we added 3 data points to the profile data to extend the profile to galactocentric radii $R<10$~kpc, hence reducing significantly the uncertainties on the fitting parameters. Available observational data for NGC~6822 extends to radii $<6.5$~kpc, and the central galaxy exponential profile dominates the halo surface brightness at radii $<4$~kpc. Hence, usable data points to fit the stellar halo are only at radii between $4$~kpc and $6.5$~kpc. Such a short span in radii is insufficient to determine the model parameters with reasonable uncertainties. Fig.~\ref{fig:ngc6822} illustrates how well we could measure the model free parameters using stellar halo data not yet available but within reach of current telescopes and observational techniques.

The top-left panel shows the surface density profile of NGC~6822 (points), the best fit exponential profile of the galaxy (solid blue line), the best fit ghostly halo model (dotted line), and the best fit model of the total (galaxy plus halo) surface density profile (solid orange line). The other panels show the 68\% (red line), and 95.4\% (blue line) confidence contour plots of the fitting parameters $\beta$, $\epsilon_0$ and $M_{halo}$. Each panel shows the marginalized 2-dimensional projections of the 3-dimensional parameter space, to highlight the uncertainties and the degeneracy of the parameters. Here we have kept the values of the fitting parameters of the exponential galaxy profile ($\Sigma_0^{gal}$ and $R_{exp}^{gal}$) fixed at the maximum likelihood value, and assumed $M_{cut}=10^6$~\msun.
It is clear from Fig.~\ref{fig:ngc6822} that, even assuming the prior $M_{cut}=10^6$~\msun, neglecting the uncertainties related to the subtraction from the data of the exponential profile of the central galaxy, and observing the profile to approximately double the radius where the central galaxy exponential profile dominates the surface brightness ($\sim 10$~kpc), the remaining three free parameters in the ghostly halo model ($\beta$, $\epsilon_0$, and $M_{halo}$) are highly degenerate with each other. If also the halo mass is assumed as a prior, both $\beta$ and $\epsilon_0$ can be determined accurately for the (artificially improved) data for NGC~6822. Hence, assuming a reasonable range for the dark matter halo mass $M_{host}$ of the dwarf galaxy, good data on the extended surface density profile of the stars can be used to constrain the star formation efficiency, $f_*$, in primordial galaxies at $z=7$. The converse is also possible: if we have a reliable estimate of either $\beta$ or $\epsilon_0$, the mass of the dark matter halo can be inferred from the model. 

As illustrated above, the data currently available for the six dwarf galaxies does not extend to sufficiently large radii beyond the central galaxy half-light radius to infer $\epsilon_0$ and $\beta$ (hence, $f_*$) with sufficient confidence using only one dwarf galaxy. There are too many free parameters for the too few data points to fit the extended stellar haloes, hence the results using a single dwarf galaxy show large uncertainties and degeneracy. However, combining the data for the six dwarf galaxies in our data set produces meaningful results for a simple reason: Each dwarf galaxy has a different halo mass, $M_{halo}$ and different $\Sigma_{0}^{gal}$, $R_{exp}^{gal}$. However the other free parameters describing the star formation efficiency before reionization ($\beta$, $\epsilon_0$ and $M_{cut}$) have the same values for all the six dwarf galaxies, increasing the constraints on these parameters of interest and also constraining the halo masses of the dwarfs relative to each other.

Fig.~\ref{fig:results} shows that combining the data for the six dwarf galaxies in our sample allows us to constrain $f_*$ before reionization with current data, if we make some assumptions on the halo masses of the six dwarf galaxies. The top left panel shows the observed surface density profiles of the stars in six isolated dwarf galaxies in the Local Group (see the legend; data points from KR19) and fits to the exponential galaxy profile (blue solid lines) and the ghostly halo model in this work (colored solid lines). The top right panel shows 68\% and 95\% confidence contour plots for the fitting parameters $\beta$ and $\epsilon_0$, assumed to be the same for all the dwarf galaxies.
The bottom left panel shows the stellar mass $M_*$ of pre-reionization dwarfs (or UF dwarfs) as a function of their dark matter halo masses at $z=7$. Each solid line refers to a different isolated dwarf galaxy considered in this study (see legend). The symbols show the stellar masses and the halo masses at $z=7$ of Leo~T and Leo~A. For simplicity we have assumed that the maximum dark matter masses, $M_{dm}(max)$, for Leo~T and Leo~A are equal to their fiducial values at $z=0$ (see Table~\ref{tab:fits2}) and we used equation~(\ref{eq:Mmax}) for an estimate of their $z=7$ dark matter masses. Finally, the bottom right panel shows the stellar mass as a function of the maximum dark matter halo mass of the satellites, obtained from the $z=7$ dark matter masses using equation~(\ref{eq:Mmax}). For comparison, the dotted line shows published results on $M_*$ using complementary methods based on halo-matching \citet{Behroozi2013}, and the dash dotted line (with shaded area for $1\sigma$ and $2\sigma$ uncertainties) refer to a similar but more recent study from the DES team \citep{Nadleretal:2020}.

\begin{table}
	\caption{Table of fiducial halo masses at $z=0$ adopted in this work.}
	\label{tab:fits2}
	\begin{tabular}{lccc} 
		\hline
		Dwarf Name & \multicolumn{3}{c}{Halo mass: $M_{halo}[M_{\odot}]$} \\
		 & Fiducial & Range & KR19 \\
		\hline
		Leo~T  & $3.0 \times 10^8$    & $(1.5-6) \times 10^8$ & $2.4\times 10^8$ \\
		Leo~A  & $3.0 \times 10^9$    & $(1.5-6) \times 10^9$ & $1.6\times 10^9$ \\
		WLM    & $1.0 \times 10^{10}$ & $(0.5-2) \times 10^{10}$ & $0.9\times 10^{10}$ \\
		IC~1613 & $1.2 \times 10^{10}$ & $(0.6-2.4) \times 10^{10}$ & $0.7\times 10^{10}$ \\
		IC~10   & $1.3 \times 10^{10}$ & $(0.65-2.6) \times 10^{10}$ & $1.4\times 10^{10}$ \\
		NGC~6822& $1.5 \times 10^{10}$ & $(0.75-3) \times 10^{10}$ & $1.5\times 10^{10}$ \\
		\hline
\end{tabular}
\end{table}

The results shown in Fig.~\ref{fig:results} are obtained with the following assumptions:
\begin{enumerate}
\item We adopted the following priors to reduce the degeneracy of the results: a) dark matter halo masses at $z=0$ for the six dwarfs as in Table~\ref{tab:fits2}; b) $M_{cut}=10^6$~M$_\odot$.
\item We choose the pivot mass $M_{0}$ in the equation for $f_*$ (equation~\ref{eq:fstar}) such that $\epsilon_0$ is independent of $\beta$ (this can only be done with a prior on $M_{halo}$). We therefore constrain $\epsilon_0$ at a given mass for each of the observed dwarf data. When improved observed stellar halo profiles become available our method will also be able to constrain the parameters that are either too uncertain (like $\beta$) or set here as prior (like the mass of the dark matter halo).
\item Here the prior on the halo masses are similar to fiducial values obtained by KR19, but slightly adjusted with respect to each other to obtain the same $\epsilon_0$ and $\beta$. However, the final result depends systematically on the overall mass scale assumed for the dark matter haloes of the six dwarf galaxies. 
\end{enumerate}

\begin{figure*}
\centering
\includegraphics[width=0.49\textwidth]{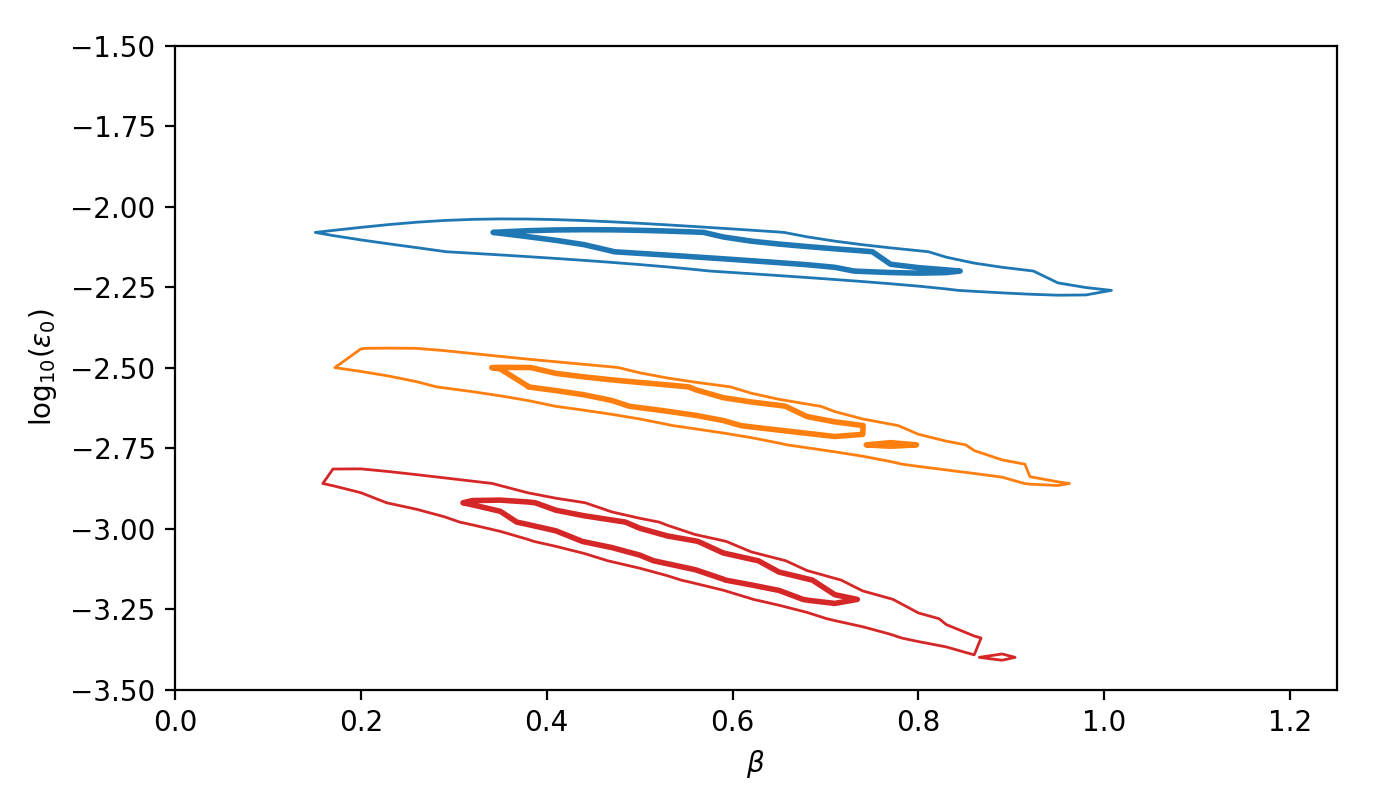}
\includegraphics[width=0.49\textwidth]{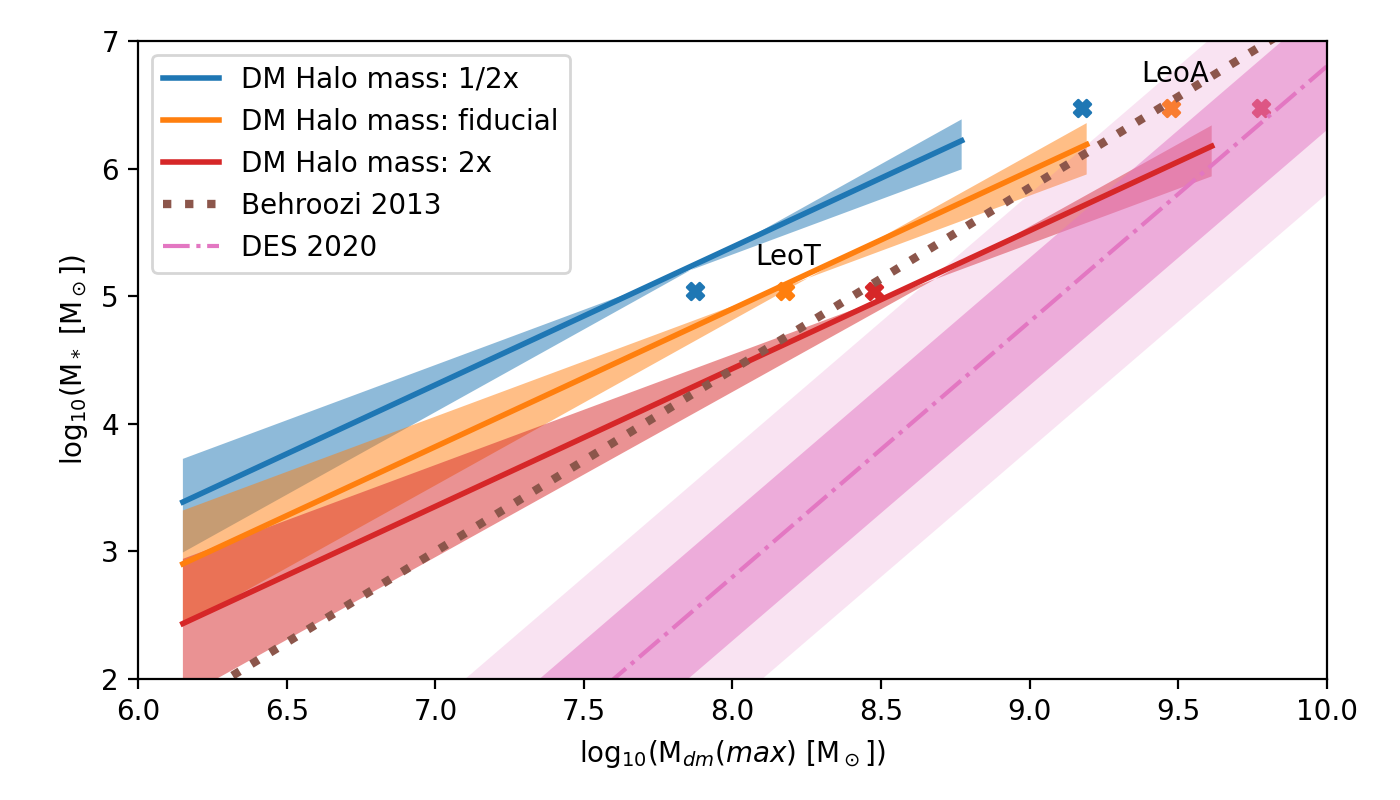}
\caption{{\it (Left).} The orange contours show the 68\% ($1\sigma$, thick lines) and 95\% ($2\sigma$, thin lines) confidence regions for the parameters $\epsilon_0$ and $\beta$ for the fiducial priors on the dark matter halo masses of the 6 dwarf galaxies in this study. The contours in blue show the same confidence region assuming a prior on the halo masses 50\% smaller than the fiducial case, and the red contour double the masses of the fiducial case.
{\it (Right).} Same as the bottom right panel in Fig.~\ref{fig:results}, but showing the uncertainty in $M_*$ due to the prior on dark matter halo masses (see the legend) and the $1\sigma$ uncertainty on the slope $\beta$.}
\label{fig:results1}
\end{figure*}

In Fig.~\ref{fig:results1} we assess the uncertainty on the fitting parameter as a result of the assumed priors on the dark matter halo masses and the uncertainty on $\beta$. The left panel shows the $1\sigma$ and $2\sigma$ confidence contours on the $\beta-\epsilon_0$ plane for different priors on the halo masses. The fiducial prior on the dark matter halo masses (see Table~\ref{tab:fits2}) is shown by the middle (orange) contour ellipses; the effect of doubling the halo masses is shown by the lower (red) ellipses and reducing the halo masses by a factor of 2 is shown by the upper (blue) ellipses. The figure shows that for a fixed prior on the halo masses the value of $f_*$ at $M=10^7$~M$_\odot$ ($\epsilon_0$) is well constrained. On the other hand, the slope $\beta$ is poorly constrained and can vary between $0.3$ and $0.8$ within the 68\% confidence contour plot.  Changing the prior on the halo masses has a direct effect on $\epsilon_0$ and produces larger values for lower halo masses.
The right panel is the same as the bottom right panel in Fig.~\ref{fig:results}, but shows the uncertainty in $M_*$ due to the prior on dark matter halo masses (see the legend) and the $1\sigma$ uncertainty on the slope $\beta$.

\section{Discussion}\label{sec:disc}

\begin{figure*}
\centering
\includegraphics[width=0.99\textwidth]{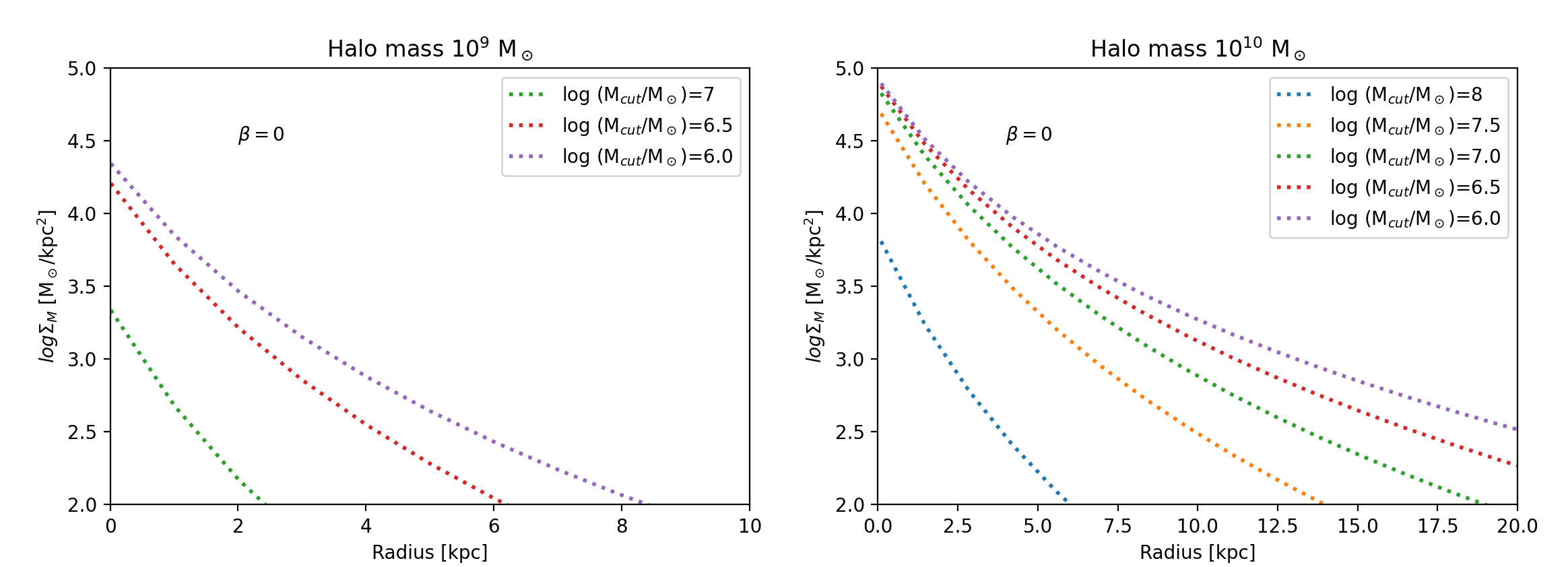}
\caption{{\it (Left.)} Model predictions for the ghostly stellar halo surface density profile as a function of radius assuming $\beta=0$, $\epsilon_0=3\times 10^{-3}$ (at $M_0=10^8$~M$_\odot$) and a mass of the dark matter halo of $M_{halo}=10^9$~M$_\odot$. Each line shows the effect of changing the cutoff mass M$_{cut}$ as shown in the legend. For a cutoff mass larger than $\sim 10^{-2}M_{halo}$ the stellar halo does not exist. {\it (Right.)} Same, as the left panel but for a dark matter halo mass $10^{10}$~M$_\odot$, corresponding to a dwarf galaxy similar to WLM. In this case if only atomic cooling halos ($M_{cut} \sim 10^8$~M$_\odot$ at $z\sim 10$) can form stars before reionization, WLM should not have a stellar halo.}
\label{fig:mcut}
\end{figure*}

Understanding galaxy scale feedback processes in the early universe and the formation of the first (non atomic cooling) galaxies has been extensively explored over the past 20 years, mostly using cosmological simulations \citep[\eg,][]{RIcotti2002, Ricotti2010}. In these simulations even low mass halos with virial temperature below the atomic cooling limit can be luminous, and indeed it has been proposed that the UF dwarfs can be explained in these models \citep{RicottiG:05, BovillR2009}. However, several authors have pushed forward models in which only atomic cooling halos can form stars before reionization  \citep[\eg,][]{BenitezFrenk:2020}. This work offers a further test to distinguish between these two scenarios,  as illustrated in Figure~\ref{fig:mcut}. The figure shows our model predictions for the ghostly stellar halo surface density profile as a function of radius for a halo with a dark matter mass similar to Leo~A ($M_{dm,halo}=10^9$~M$_\odot$, left panel) and one similar to WLM ($M_{dm,halo}=10^{10}$~M$_\odot$, right panel). Each line shows the effect of changing the cutoff mass M$_{cut}$ as shown in the legend. For a cutoff mass larger than $\sim 10^{-2}M_{dm, halo}$ the stellar halo does not exist. For the case of WLM, if only atomic cooling halos ($M_{cut} \sim 10^8$~M$_\odot$ at $z\sim 10$) can form stars before reionization, the galaxy should not have a stellar halo or be too faint to detect \citep{Deasonatal:2022}.

One of the assumptions in our model is that 
dwarf galaxy stellar halos are comprised of pre-reionization fossil galaxies, hence they do not experience major mergers after reionization.
This was supported in Figure~\ref{fig:mergerhist}, showing that the dark matter halos of isolated dwarf galaxies typically experience slow growth after the time of virialization at high redshift. Therefore major mergers mostly happen at high-z when the galaxy grows rapidly and virializes, while at late times the slow accretion rate is driven by minor mergers. We found that all the 9 isolated halos that we have simulated with masses $<10^{10}$~M$_\odot$, do not experience major mergers and accrete only halos that do not form stars after reionization (with $T_{vir}<T_{IGM}$). However, dwarf galaxies more massive than this mass threshold are more likely to accrete satellites that continue forming stars after reionization. In this case some of the stars in the stellar halo should be younger and more metal rich.
We expect, however, that these younger stars are deposited mostly in the inner parts of the stellar halo because the merging halo is massive. We have already discussed how we should strive to only use the outer parts of the halo to avoid possible contamination from stars from the central galaxy ejected by mergers and feedback effects.
In practice observations of ghostly halos can only be done by star counts using a CMD to select the RG branch for an old and metal poor stellar population at the distance of the galaxy. Therefore observations already select against contaminants such as the ones discussed above.
Eventually our method to constrain $f_*$ will fail in more massive dwarfs or even Milky-Way mass galaxies because the stellar halo is composed of stars that formed at all redshifts. Therefore such halos constrain $f_*$ not at the time of reionization, but at the time just before most of the mergers occur (when the merging halos reach $v_{max}$). The outer parts of any stellar halo, however, should be dominated by stars from accreted "fossil galaxies.

The statistical uncertainty on our result can be greatly improved by increasing the accuracy and the outer radii over which the surface brightness is measured for the six dwarf galaxies in our sample. In addition, it is possible to add more dwarf galaxies to our sample as discussed in \cite{KangR:2019}.
Obtaining data further from the central galaxy exponential profile is important to rule out contamination of the halo profile by stars ejected from the central galaxy by dynamical processes internal to the galaxy (\eg, feedback) or external to the galaxy (\eg, major galaxy mergers), and hence reduce the possibility of systematic errors. 
The impact of mergers and feedback effects is expected to produce the strongest effects on the halo profile in the inner regions of the halo, that are currently used to estimate $f_*$. We expect that this may lead to an overestimate of the halo stars and therefore $f_*$.
However, as explained above, observations select old and metal poor stars in the halo, reducing the potential effect of contaminants such as the ones discussed above.

The space for improvement enabled by future observations is vast and exciting. In November 2021 with one night of observing time on the Discovery Channel Telescope at Lowell observatory, we have obtaining photometry in V and I for two fields offset from WLM to study the outer parts of the ghostly halo profile that currently lacks data. We reached 26 and 25 mag in V and I respectively and we are rather confident we have detected the stellar halo, although follow up observations in other directions around WLM are needed to draw robust conclusions. In addition to WLM, there are many other targets to pursue with sufficiently deep observations: the outer parts of the six dwarfs in this study, but also several other dwarfs which so far do not show any evidence of an extended stellar halos but we have suggested in \cite{KangR:2019} as promising targets.

Our method to constrain $f_*$ relies on some assumptions that may produce physically-motivated discrepancies with respect to methods based on halo-matching between the halo mass and the luminosity of dwarf galaxies \citep[\eg,][]{Behroozi2013, Nadleretal:2020}. Discovering such a discrepancy would be interesting on its own merit. For instance, in our study we have not considered the contribution of globular clusters to the build up of the stellar halo. More precisely, although we have explored the case in which the stellar spheroid at the center of each halo is rather compact, we did not consider the case of multiple compact stellar system in each merging halo (\ie, a system of GCs). 

Perhaps more interesting is the possibility that our method captures the full stellar content of haloes at the time of formation while methods based on halo matching of UF dwarfs capture only the stars presently bound by the dark matter halo. Today's stellar mass in UF dwarf galaxies may be significantly smaller than at formation not only because of tidal stripping, but also because stars with high velocity dispersion may evaporate out of small mass haloes and disperse in the IGM or ICM. A mechanism in which this process is likely to happen in UF dwarf galaxies was proposed in \cite{RicottiPG2016}. In that work, the origin of UF dwarfs was linked to the evaporation or destruction in them of compact star clusters with initial velocity dispersion typical of proto-GCs (half-light radii of few pcs and $\sigma_* \sim 30 -40$ km/s), much greater than the circular velocity of the minihaloes in which they formed ($\sim 5-15$ km/s). As these proto-GCs expand and disperse, most of the stars remain bound by the gravitational potential dark matter minihalo in which they form, producing an UF dwarf (the velocity dispersion of the star cluster decreases as it expands). But it is also possible that a significant fraction of the stars escape the haloes (depending on the initial size and velocity dispersion of the star cluster with respect to $r_{max}$ and $v_{max}$ of the minihalo).    
These stars would escape the minihaloes in which they form but likely they will still contribute to the stellar halo of the host galaxy they eventually merge with. Therefore, our method would systematically predict higher stellar masses for a given halo mass than methods based on halo-matching.

Both in this work and in KR19 we made an assumption of one-to-one relationship between halo mass and stellar mass in the pre-reionization fossil galaxies at $z=7$. However, simulations of galaxy formation show that in the mass range $10^6-10^8$~M$_\odot$, the scatter around a mean star formation efficiency is very large \citep[{\it e.g.},][]{RicottiGS2002b, Fitts:2017, Munshi:2017}. The effects of this stochasticity in the mass-to-light ratio in the first galaxies needs to be included in future models to check whether it has an effect on the scale radius of ghostly haloes.

\section{Summary and Conclusions}\label{sec:conc}

We have used N-body simulations to understand and model the formation of stellar haloes in dwarf galaxies. When applied to observations of dwarf galaxies with today's masses $< 10^{10}$~\msun, the model can be used to infer the star formation efficiency of fossil dwarf galaxies (UF dwarfs) forming the bulk of their stars before the epoch of reionization at $z \sim 7$. 

The reason is easy to understand: if a stellar halo is found in sufficiently small mass dwarfs, the whole stellar halo is composed of tidal debris of fossil galaxies, {\it i.e.,} galaxies that formed most of their stars before the epoch of reionization. These stellar haloes made of only old stars have been referred to as "ghostly" stellar haloes \citep{BovillR2011b}. Therefore the detection and characterization of ghostly stellar haloes around isolated dwarf galaxies is a sensitive test of the efficiency of star formation in fossil galaxies. Clearly the properties of ghostly haloes are tightly connected to the properties of the population of UF dwarf galaxies found around the Milky Way and M31. This elusive galaxy population still needs to be fully uncovered and understood and is one of the most powerful probes of the physics of galaxy formation in the early universe and the epoch of reionization.

We have derived an empirical model of ghostly stellar haloes using a set of cosmological N-body simulations including dark matter haloes and stars. The basic idea of the model is physically motivated: the profile of the stellar halo can be understood as the sum of exponential profiles with scale radii determined by the typical masses of the accreting subhaloes building up the galaxy. Smaller mass subhaloes deposit most of their stars in the outer parts of the halo profile and vice versa. We apply the model to interpret observational data for six isolated dwarf galaxies in the local group (Leo~T, Leo~A, WLM, IC~10, IC~1613, NGC~6822) with masses ranging from $\sim 10^8$~\msun\ to $\sim 10^{10}$~\msun\ and infer the star formation efficiency, $f_*$, at $z \sim 7$ in dark matter minihaloes with masses between $10^6$~\msun\ to $10^8$~\msun.
The following is a summary of our findings:
\begin{enumerate}
\item  Current data on each single dwarf galaxy is not good enough for constraining all the model parameters and infer $f_*(M)$ at $z=7$ with reasonably small errorbars. This is due to the difficulty of separating the exponential galaxy profile from the halo profile due to the paucity of data at large distance from the galaxy and the large errorbars of the existing distant data points.
\item If we use the data of only one dwarf galaxy, the value $\epsilon_0$ is highly degenerate with $M_{halo}$(z=0) ($\epsilon_0$ is roughly inversely proportional to $M_{halo}$(z=0)). Hence, even if we acquire exquisite data extending beyond what is currently available in published observations, a prior on the halo mass of the dwarf galaxy is necessary to derive $f_*(M)$ in pre-reionization dwarf galaxies.
\item
Although each galaxy has its own exponential profile parameters, $f_*(M)$ for the accreted satellites (\ie, $\epsilon_0$ and $\beta$) are the same for all six galaxies. We therefore can combine the likelihood ellipses of each galaxy to find the values that maximize the combined likelihood. This allows us to constrain $f_*(M)$ with current data using a prior on the halo masses. Better data that extends to larger galactocentric distance will allow us to constrain both $f_*(M)$ at $z=7$ and the dark matter halo masses of the Local Group dwarfs.
\item For all galaxies but the smallest ones (Leo~A and Leo~T), it is not possible to constrain the cutoff mass for $f_*$ as small mass satellites in the range $10^6-10^7$~\msun\ affect the outer parts of the stellar haloes for which data is not currently available. However, if the existence of extended stellar haloes around Leo~T and Leo~A are confirmed, a cutoff of the luminosity function at $M_{cut} \ge 10^7$~\msun\ would be incompatible with the data, as the stellar halo would not exist with a cutoff at this mass. This is the main reason for our assumption of $M_{cut}=10^6$~\msun.
\end{enumerate}
Our results show that $f_*$ is in agreement with previous works \citep{Behroozi2013, KangR:2019, Nadleretal:2020} but the statistical uncertainty on the result is still large because the data on the observed dwarf galaxies does not extend sufficiently far from the scale radius of the central galaxy exponential profile. We conclude that this new method to constrain the star formation efficiency in the first galaxies before reionization, which was first introduced in \cite{KangR:2019} and refined in the present work, provides a new powerful tool to test the consistency of models of galaxy formation. It also offers a compelling theoretical motivation to collect more and better observational data on the extended stellar haloes around dwarf galaxies. The stellar haloes in small mass dwarf galaxies and the outer parts of more massive dwarf galaxies contain stars from the smallest satellite building blocks and therefore are most sensitive to determining the star formation efficiency in the first galaxies.

In addition to new observational data, there are several further questions that need to be explored in more detail with respect to the modeling aspect. i) Within the stellar halo we expect to observe surviving satellites with luminosities comparable to Milky Way's UF dwarfs, but most satellite are in the outer parts of the halo and therefore it will be time-consuming to obtain deep enough observations around a large area in the sky to detect them. A preliminary analysis shows that the number of surviving satellites scales with the prominence of the stellar halo, as expected. In Halo D assuming $\beta=1$ and $M_{cut}=10^6$~M$_\odot$ we observed 10 satellites within a galactocentric distance of 50 kpc (and 4 within 20 kpc). The number is reduced to 4 (0 within 20 kpc) assuming $M_{cut}=10^7$~M$_\odot$, while is much greater in models with $\beta=0$: 25 (8 within 20 kpc) for $M_{cut}=10^6$~M$_\odot$, and 5 (2 within 20 kpc) for $M_{cut}=10^7$~M$_\odot$.  ii) A characterization of the stellar haloes triaxiality and anisotropicity due to tidal shocks and streams will be required to better compare observations to the models, especially if the stellar halo is observed using only sparse covering on the sky. iii) Finally, we point out that the simulations and our halo model naturally predict a metallicity gradient of the stellar halo, because the stars from the lowest mass satellites, that are more metal poor, are deposited preferentially at larger galactocentric radii. We plan to explore quantitatively these issues in future work.

\section*{Acknowledgements}
We thank the anonymous referee for the prompt and insightful report that helped us improve the quality of this paper. We thank Dr. Volker Springel for sharing the {\sevensize GADGET~3} code with us. Visualisations in Fig.~\ref{fig:dm_portraits} and Fig.~\ref{fig:images} were created with {\sevensize SPLASH} \citep{splash} using particle smoothing lengths calculated by {\sevensize HOP} \citep{hop}. All the simulations were performed with the Deepthought2 cluster operated by the University of Maryland (http://hpcc.umd.edu). MR acknowledges the support by NASA grant 80NSSC18K0527. Basic research in astronomy at the Naval Research Laboratory is funded by 6.1 Base funding.

\section*{Data Availability}
The data underlying this article will be shared on reasonable request to the corresponding author.




\bibliographystyle{mnras}
\bibliography{ghostly_sim} 





\bsp	
\label{lastpage}
\end{document}